\documentclass[apj]{emulateapj}

\shorttitle{GALAXY SIZE EVOLUTION}

\shortauthors{I. TRUJILLO ET AL}

\slugcomment{To appear in ApJ.}

\begin{document}

\title{The  size evolution of galaxies since
z$\sim$3: combining SDSS, GEMS and FIRES\altaffilmark{1}}

\altaffiltext{1}{Based on observations collected at the European Southern
Observatory, Paranal, Chile (ESO LP 164.O--0612). Also, based on observations
with the NASA/ESA $Hubble$ $Space$ $Telescope$, obtained at the Space Telescope
Science Institute, which is operated by AURA Inc, under NASA contract NAS
5--26555.}

\author{Ignacio Trujillo\altaffilmark{2,3}, Natascha M. F\"orster
Schreiber\altaffilmark{4}, Gregory Rudnick\altaffilmark{5}, Marco
Barden\altaffilmark{2},   Marijn Franx\altaffilmark{6},  Hans--Walter
Rix\altaffilmark{2}, J. A. R. Caldwell\altaffilmark{7}, Daniel H.
McIntosh\altaffilmark{8}, Sune Toft\altaffilmark{9}, Boris H\"au\ss
ler\altaffilmark{2}, Andrew Zirm\altaffilmark{6}, Pieter G. van
Dokkum\altaffilmark{9}, Ivo Labb\'e\altaffilmark{10}, Alan
Moorwood\altaffilmark{11}, Huub R\"ottgering\altaffilmark{6}, Arjen van der
Wel\altaffilmark{6}, Paul van der Werf\altaffilmark{6}, Lottie van
Starkenburg\altaffilmark{6}}

\altaffiltext{2}{Max-Planck-Institut f\"ur Astronomie, K\"onigstuhl 17, 69117
Heidelberg, Germany}
\altaffiltext{3}{School of Physics \& Astronomy, University of Nottingham,
University Park, Nottingham, NG7 2RD, UK}
\altaffiltext{4}{Max-Planck-Institut f\"ur extraterrestrische Physik,
Giessenbachstrasse, D-85748, Garching, Germany}
\altaffiltext{5}{NOAO, 950 N. Cherry Av. Tucson AZ 85719}
\altaffiltext{6}{Leiden Observatory, P.O. Box 9513, NL--2300 RA, Leiden, The Netherlands}
\altaffiltext{7}{University of Texas, McDonald Observatory, Fort Davis, TX 79734}
\altaffiltext{8}{Astronomy Department, University of Massachusetts, 710 N.
Pleasant St., Amherst, MA 01003}
\altaffiltext{9}{Department of Astronomy, Yale University, P.O. Box 208101,
New Haven, CT 06520-8101}
\altaffiltext{10}{Carnegie Observatories, 813 Santa Barbara Street,  Pasadena, CA 91101}
\altaffiltext{11}{European Southern Observatory, D--85748, Garching, Germany}

\begin{abstract}

We present  the evolution of the luminosity--size and stellar mass--size
relations of luminous
(L$_V$$\gtrsim$3.4$\times$10$^{10}$h$_{70}$$^{-2}$L$_\odot$) and of massive
(M$_\star$$\gtrsim$3$\times$10$^{10}$h$_{70}$$^{-2}$M$_\odot$) galaxies in the
last $\sim$11 Gyr. We use very deep near--infrared images of the Hubble Deep
Field--South and the MS1054-03 field in the J$_s$, H and K$_s$ bands from FIRES
to retrieve the sizes in the optical rest--frame for galaxies with z$>$1. We
combine our results with those from GEMS  at 0.2$<$z$<$1 and SDSS at z$\sim$0.1
to achieve a comprehensive picture of the optical rest--frame size evolution 
from z=0 to z=3. Galaxies are differentiated according to their light
concentration using the S\'ersic index $n$. For less concentrated  objects, the
galaxies  at a given luminosity were typically $\sim$3$\pm$0.5 ($\pm$2 $\sigma$)
times smaller at z$\sim$2.5 than those we see today. The stellar mass--size
relation has  evolved  less: the mean size at a given stellar mass was
$\sim$2$\pm$0.5 times smaller at z$\sim$2.5, evolving proportional to
(1+z)$^{-0.40\pm0.06}$. Simple scaling relations between dark matter halos and
baryons in a hierarchical cosmogony predict a stronger (although consistent
within the error bars) than observed evolution of the stellar mass--size
relation.  The observed luminosity--size evolution out to z$\sim$2.5 matches
well recent infall model predictions for Milky--Way type objects. For low-n
galaxies, the evolution of the stellar mass--size relation would follow
naturally if the individual galaxies grow inside--out. For highly concentrated
objects, the situation is as follows:  at a given luminosity, these galaxies
were $\sim$2.7$\pm$1.1 times smaller at z$\sim$2.5 (or put differently, were
typically $\sim$2.2$\pm$0.7 mag brighter at a given size than they are today),
and at a given stellar mass the size has  evolved proportional to
(1+z)$^{-0.45\pm0.10}$.

\end{abstract}

\keywords{galaxies: fundamental parameters,
galaxies: evolution, galaxies: high redshift, galaxies: structure}




\section{INTRODUCTION}

Over the last few decades (starting with Fall \& Efstathiou 1980 and Fall 1983)
there has been a substantial effort towards understanding, theoretically and
through observations, how galaxies have reached their current sizes over cosmic
time. The answer to this question plays a key role in our understanding of
galaxy formation and evolution.

Several approaches have been tried to make specific predictions about how  sizes
of galaxies (particularly the disk galaxies) evolves with redshift:
semi--analytical hierarchical models, direct numerical simulations  and  infall
models.

The  semi--analytical hierarchical model assumes simple scaling relationships
between the properties of the galaxy disks and the halos in which they reside
(Lacey et al. 1993; Kauffmann \& Charlot 1994; Dalcanton et al. 1997; Mo, Mao \&
White 1998; Sommerville \& Primack 1999; van den Bosch 2000; Cole et al. 2000;
Naab \& Ostriker 2006).
According to this picture, galaxy disks are formed from gas with some initial
angular momentum that cools and contracts in dark matter halos. The mass and the
angular momentum that settle in the disk are some fixed fractions of the mass
and the angular momentum of the halo respectively. The mass and size of the
halos are tightly linked to the density of the universe at the time the halos
were formed; consequently, halos formed at high--z are expected to be much
denser than halos formed at lower z. Under the assumption that the fractions of
disk mass and angular momentum in the disk relative to the halo, together with
the spin parameter of the halo do not vary with redshift,  Mo et al. suggest the
following redshift scaling for the size of the baryonic disk at their
$formation$ redshift:  R $\propto$ H$^{-1}$(z) at a fixed circular halo velocity
or R $\propto$ H$^{-2/3}$(z) at a fixed halo mass, where H(z) is the Hubble
constant at a given z: H(z)=H$_0$[$\Omega_m$(1+z)$^3$+$\Omega_\Lambda$]$^{1/2}$
in a flat Universe. 


 High--resolution N--body/gas-dynamical simulations (Navarro \&
Steinmetz 2000; Brook et al. 2006) find that the above picture is too simplistic; e.g. large
systematic variations in the fraction of baryons that collapse to form galaxies
are observed and angular momentum conservation may not hold. Moreover, the
explanation of the observed local size--mass relation within this hierarchical
context (Shen et al. 2003) requires that the above scaling between the dark
matter and baryons is broken and instead that the fraction of baryons in the disk
is a function of the halo mass.  This is also predicted by standard feedback
models based on galactic winds.

The infall model approach (Cay\'on, Silk \& Charlot 1996; Bouwens, Cay\'on \&
Silk 1997) examines a number of local disk galaxies in great detail and uses
detailed models of their observed properties, e.g. gas profiles, stellar
profiles, metallicity profiles, current star formation rate (SFR), and
age--metallicity relationships, to infer how galaxies might have evolved from
high redshift. This approach uses the local universe as a reference and
consequently does not explain why the local galaxy population is as it is. The
main ingredients of these models  are: a) that the SFR is determined at each
radius and time from the local gas density according to a Schmidt--type law, and
b) that metal--free gas infalls with certain time-scale. Using the Milky Way as
reference, Bouwens \& Silk (2002) provide the following size scaling
relationship with redshift: R(z)/R(0)=1-0.27z.

 In the case of spheroid--dominated galaxies, they  are expected to form  from
the merging of smaller systems (White \& Frenk 1991) and consequently to have a
different size evolution than disk--dominated systems. The old stellar
populations found in nearby ellipticals make it unlikely that these galaxies
were the remnant of a merger between two similar spirals drawn from the observed
local population. In fact, Khochfar \& Burkert (2003) have shown that
dissipationless mergers of early--type galaxies may dominate the formation of
the nowadays high--mass early--type galaxies. In addition, there is some
observational (van Dokkum 2005; Tran et al. 2005; Bell et al. 2006) and
theoretical (Naab, Khochfar \& Burkert 2006; Boylan--Kolchin et al. 2006)
evidence pointing towards the merger of red galaxies as the potential formation
mechanisms for the spheroid population. Shen et al. (2003) have shown that the
present--day stellar mass--size relation for early--type galaxies follows
R$\propto$M$^{0.56}$. Shen et al. indicate that the present--day relation is
consistent with a model where early--type galaxies are the remnants of repeated
mergers where the progenitors have properties similar to those of faint
ellipticals. According to their model, the size of the remnant increases after
each merger. In this context, we would expect that early--type galaxies that
have undergone a major merger were  larger in size than galaxies of the same
mass that have not suffered such a process.


 Detailed modeling of the merger histories of galaxies in the cold dark matter
scenario suggests that the last major merging event is typically around redshift
unity (Kauffmann \& Haehnelt 2000). Consequently, we would expect that the sizes
of early--type galaxies at z$>$1 were, in general,  smaller than the local
counterparts. An analysis of the evolution  of the stellar mass--size relation
at high--z of these objects can constrain the above scenario of merging
formation. 

Historically, the monolithic collapse scenario (Eggen, Lynden--Bell \& Sandage
1962; Larson 1975) envisioned that all spheroidal galaxies formed very early via
a rapid collapse of the gas at high redshift. In this picture, E/S0s  would 
already be in place at high--z and we would expect then that the changes in the
observed properties of early--type galaxies over time were due to simple passive
fading of their stellar populations. The more modern version of this scenario
(e.g. Chiosi \& Carraro 2002; Merlin \& Chiosi 2006) envisions that  massive
ellipticals also formed hierarchically, but at quite high redshift.

The evolution of individual galaxies is not directly observable. However,
look--back studies can provide extensive information on how the population
properties of galaxies have changed with cosmic epoch. Early studies (Smail et
al. 1995; Casertano et al. 1995; Roche et al. 1998) showed that galaxies at a
given luminosity  were smaller in the past.  However, it was not until the
application of the Lyman-break technique (Steidel et al.  1996) that the study
of a large number of galaxies at high--z  was possible. This technique is
especially efficient at selecting star--forming galaxies at z$>$2. Sizes  have
been measured for these Lyman Break Galaxies (LBGs) (Giavalisco, Steidel \&
Macchetto 1996; Lowenthal et al. 1997; Ferguson et al. 2004), but using optical
filters, i.e. measuring their sizes in the rest--frame ultra-violet (UV) region
of their spectra. At these wavelengths the LBGs appear compact
(r$\sim$0.''2--0.''3, $\sim$1.5--2.5 h$_{70}$$^{-1}$ kpc). However, there is
some evidence that the LBG morphology depends very little on the wavelength,
remaining essentially unchanged from the far--UV to the optical window
(Giavalisco 2002; Papovich et al. 2005).

As a result of the dearth of very deep near--infrared (NIR) images, most of the
studies using the rest--frame optical have been limited in redshift up to z
$\sim$1 (Schade et al. 1996; Lilly et al. 1998; Simard et al. 1999; Ravindranath
et al. 2004; Trujillo \& Aguerri 2004; McIntosh et al. 2005; Barden et al.
2005). To properly compare with local optically selected samples and to trace
the size evolution in a consistent fashion at z$>$1 one needs to use very deep
NIR data. Consequentially any observed size evolution would then reflect true
evolutionary changes not subject to the changing appearance of galaxies in
different bandpasses. Moreover, it seems now clear that rest-frame UV selected
samples do not provide a complete census of the galaxy population at high--z
(e.g. Franx et al. 2003; van Dokkum et al. 2003; Daddi et al. 2004) and, in
particular, a substantial population of red objects are missing from purely
rest-frame UV selected surveys.

In addition to the use of rest--frame optical sizes, it would be of
great help to facilitate a direct comparison with the theoretical
expectations if the size evolution could be measured at a given mass
rather than a given luminosity. Using circular velocity
measurements to estimate galaxy masses at high--z is 
difficult and  few objects have been analyzed (see e.g. Vogt
et al. 1996; 1997; Boehm \& Ziegler 2006; Erb et al. 2006). An alternative approach is to estimate the
stellar masses from their rest--frame colors and spectral energy
distributions (SEDs).

With the above ideas in mind we performed an exploratory work (Trujillo et al. 
2004) to probe the evolution of the luminosity--size and stellar mass--size
relations of the galaxies out to z$\sim$3. That work used very deep NIR images
of the Hubble Deep Field--South (HDF--S) from the Faint Infrared Extragalactic
Survey (FIRES; Franx et al. 2000). We found that the rest--frame V--band sizes
of luminous galaxies ($<$L$_V$$>$$\sim$4$\times$10$^{10}$
h$_{70}$$^{-2}$L$_\odot$) at 2$<$z$<$3 were 3 times smaller than for equally
luminous galaxies today. In contrast, the stellar mass--size relation had
evolved relatively little: the size of galaxies more massive than
2$\times$10$^{10}$h$_{70}$$^{-2}$M$_\odot$, were $\sim$1.5 times smaller at
z$\sim$2.5\footnote{During the writing of the present paper we discovered a bug
in the code which was used to estimate the sizes in the 2004 paper. The sizes of
the smallest objects in our HDF-S sample (r$_e$$<$0.2$\arcsec$) were
overstimated. This produced a slight underestimation on the degree of evolution
in the luminosity and stellar mass size relation. This problem has been solved
in the present version.}.

In the present work we add to the above data set the results from the analysis of
the $\sim$ 4 times larger MS1054--03 FIRES field. Using both FIRES fields  we
decrease the effects  of the field--to--field variations in our results and
multiply by three the number of objects with z$>$1 in our sample. In addition, we
make a detailed comparison of our results with  those found in the Sloan Digital
Sky Survey (SDSS; York et al. 2000) at z$\sim$0.1 and in the Galaxy Evolution
from Morphology and SEDs (GEMS; Rix et al. 2004)  survey at intermediate redshift
0.2$<$z$<$1. This allows us to follow in detail the evolution of the
luminosity--size and stellar mass-size relations of the luminous galaxies over the last
$\sim$ 11 Gyr.

The structure of this paper is as follows. In Sect. 2 we describe the FIRES
data, and in Sect. 3 the size measurement technique and robustness estimations
for the FIRES data. In Sect. 4 we present the observed luminosity--size and
stellar mass--size relations and compare our results with other samples in
Sect. 5. We discuss our results in Sect. 6.

All magnitudes in this paper are given in the AB system unless otherwise stated.
  Throughout, we will assume a flat $\Lambda$--dominated cosmology
  ($\Omega_M$=0.3, $\Omega_{\Lambda}$=0.7 and $H_0$=70  km s$^{-1}$
  Mpc$^{-1}$).

\section{FIRES: DATA}

The data used here were obtained as part of FIRES (Franx et al. 2000),  a
non--proprietary NIR survey of the HDF--S and MS 1054--03 fields carried out at
the European Southern Observatory (ESO) Very Large Telescope (VLT). The data
processing and photometry are discussed in detail by Labb\'e  et al. (2003a) for
HDF--S and F\"orster Schreiber et al. (2005) for the  MS 1054--03
field\footnote{The reduced images, photometric catalogs, photometric redshift
estimates, and rest--frame luminosities are available online through the FIRES
home page at http://www.strw.leidenuniv.nl/~fires.}.

The NIR images were obtained using the VLT Infrared Spectrograph And Array
Camera (ISAAC; Moorwood et al. 1997). The HDF--S was imaged for 33.6 hr in
J$_s$, 32.3 hr in H, and 35.6 hr in K$_s$ in a single 2.$'$5 $\times$  2.$'$5
pointing covering the Hubble Space Telescope (HST) WFPC2 main field. The NIR
data were complemented with deep optical publicly available HST WFPC2 imaging in
the U$_{300}$, B$_{450}$, V$_{606}$ and I$_{814}$ bands (Casertano et al.
2000). For the MS 1054--03 field, 77 hr of ISAAC integration time was obtained
in a 5$'$ $\times$  5$'$ mosaic of four pointings. Already existing mosaics in
the WFPC2 V$_{606}$ and I$_{814}$ bands (van Dokkum et al. 2000) were used. In
addition, Bessel U, B, and V band imaging with the VLT FORS1 instrument were 
collected.

The depth (3 $\sigma$) reached was 26.8 mag in J$_s$, 26.2 mag in H,  and
26.2 mag in K$_s$ for point sources in the HDF--S. The MS 1054--03 field
surveys an area four times larger down to $\sim$ 0.7 mag brighter magnitudes.
The effective seeing in the reduced images is approximately 0.$''$47 in all
NIR bands in the HDF--S and 0.$''$49 in the MS 1054--03 field.

The sources were selected in the K$_s$ band using version 2.2.2 of the
SExtractor software (Bertin \& Arnouts 1996). For consistent photometry across
all bands, the fluxes were measured on the maps convolved to a common spatial
resolution, matching the map of poorest seeing. Colours and spectral energy
distributions used in this work are based on measurements in custom isophotal
apertures defined from the detection map. Total magnitudes in the K$_s$ band
were computed in apertures based on autoscaling apertures (Kron 1980) for
isolated sources and adapted isophotal apertures for blended sources. The
photometric uncertainties were derived empirically from simulations on the
maps.

K band selected samples ensure, for $z$$\lesssim$3 galaxies, a selection based
on flux at wavelengths redder than the rest--frame V band. This selection is
less sensitive to   unobscured star formation  than selections based in the
rest--frame UV bands. From the above K band catalogs we removed  stars if their 
spectral energy distributions (SEDs) were better fitted by a single stellar
template than by a linear combination of galaxy templates. In the HDF-S two
obviously extended objects were removed  from the star lists and in the
MS1054-03 field, 4 bright spectroscopically  identified stars were added to the
star lists.

Photometric redshifts z$_{ph}$, as well as the rest--frame optical luminosities,
were estimated by fitting a linear combination of redshifted SEDs of  galaxies of
various types (Rudnick et al. 2001, 2003). Comparison with available
spectroscopic redshifts z$_{sp}$ implies an accuracy of
$\delta$z$\equiv$$<|z_{sp}-z_{ph}|/(1+z_{sp})>$=0.074 for both fields. When
possible, spectroscopic redshifts were used.

To ensure sufficient signal--to--noise ratio for the subsequent size
determinations we selected only galaxies with K$_s$$\leq$23.5 in the HDF--S and
K$_s$$\leq$23 in the MS 1054--03 field and whose fractional exposure time in all
the filters were larger than 15\% of the maximum in each field. This leaves us
with a total sample of 171 objects in the  HDF--S and 708 in the MS 1054--03
field. In part, the large number of objects in the MS 1054--03 field is caused
by a ``foreground'' cluster at $z$=0.83. To avoid  possible contamination   in
our field galaxy analysis by cluster galaxies we select only objects with
z$\geq$1. This is particularly effective at bright magnitudes due to the high
spectroscopic completeness for cluster members. For homogeneity,  the same z cut
is used in the HDF--S in the present work.

The final number of galaxies used in this paper is 87 in the HDF--S and  175 in
the MS 1054--03 field.

The stellar mass--to--light (M/L) ratio and hence the stellar masses of the
objects are estimated by Rudnick et al (2006), using
rest--frame (B-V) color and SEDs   similar to that of Bell \& de Jong (2001).
We use the relation between color and M/L, which exists over a wide range of
monotonic star formation histories and is rather robust against the effects of
age, dust extinction, or metallicity. The largest systematic errors in the
derived stellar mass will occur for galaxies with strong ongoing bursts.

\section{FIRES: REST--FRAME SIZE ESTIMATIONS}

\label{sizemeas}

The galaxy sizes used in this paper are measured in the observed band that is
closest to the rest--frame V--band at every redshift; this means  J$_s$ for
galaxies with 1$<$z$<$1.5, H for galaxies with 1.5$<$z$<$2.6 and K$_s$ for
galaxies with 2.6$<$z$<$3.2. In addition, we have also measured the sizes of all
our galaxies in the K$_s$ band to analyze the completeness of the sample and
test the robustness of the retrieved structural parameters. The structural
properties of the galaxies are estimated from a S\'ersic (1968) r$^{1/n}$ model
convolved with the image point-spread function (PSF) using the two-dimensional
fitting code GALFIT (Peng et al. 2002). The PSF  (in all the NIR bands) is very
stable with a standard deviation in the FWHM $<$3\% throughout the explored
field of view. Best--fitting stellar parameters are summarized in  Table
\ref{psfdata}. The S\'ersic model is given by \begin{equation}
I(r)=I(0)\exp\biggl[-b_n\biggl(\frac{r}{r_e}\biggr)^{1/n}\biggr], \end{equation}
where $I(0)$ is the central intensity and $r_e$ the effective radius enclosing
half of the flux from the model light profile.  The quantity $b_n$ is a function
of the radial shape parameter $n$ -- which defines the global curvature in the
luminosity profile -- and is obtained by solving the expression $\Gamma
(2n)$$=$$2\gamma (2n,b_n)$, where $\Gamma(a)$ and $\gamma(a,x)$ are respectively
the gamma function and the incomplete gamma function (see Graham \& Driver 2005
for a recent review of the S\'ersic model).

The S\'ersic model is a  flexible parametric  description of the surface
brightness distribution of the galaxies and contains the exponential (n=1) and
de Vaucouleurs (n=4) models as particular cases. In addition, this model is
used in the structural analysis of the SDSS galaxy sample (our local comparison
sample; Blanton et al. 2003; Shen et al. 2003) and the GEMS data (our 
comparison sample for galaxies in the redshift range 0.2$<$z$<$1; Barden et al.
2005; McIntosh et al. 2005).

GALFIT convolves S\'ersic profile galaxy models with the PSF of the images and
then determines the best fit by comparing the convolved models with the science
data using a Levenberg--Marquardt algorithm to minimize the $\chi^2$ of the fit.
Neighboring galaxies were excluded from each model fit using a mask, but in the
case of closely neighboring galaxies with overlapping isophotes, the galaxies
were fitted simultaneously.

 In what follows, we refer to the ``circularized effective radius'' of the
fitted model, i. e., $r_e=a_e\sqrt{(1-\epsilon)}$, where $a_e$ is the semimajor
effective radius (directly measured in our fits) and $\epsilon$ the intrinsic
(non--seeing affected) projected ellipticity of the galaxy. The results of our
fitting are shown in Table \ref{tabdata} for the MS1054--03 data. For
consistency, the HDF--S data estimated using GALFIT are also provided here
(Table \ref{tabdatasur}). 


\subsection{Structural Parameter  Estimates}

\subsubsection{Simulations}
\label{simparameters}

The results presented in this paper rely on our ability to measure accurate
structural parameters. To gauge the accuracy of our parameter determination we
have created 1000 artificial galaxies uniformly generated at random in the
following ranges: 18$\leq$ K$_s$(AB)$\leq$24, 0.$''$03$\leq$r$_e$$\leq$3$''$,
0.5$\leq$n$\leq$8 and 0$\leq$$\epsilon$$\leq$0.8. To simulate the real
conditions of our observations, we add a background sky image (free of sources)
taken from a piece of the MS1054 field  image in the K$_s$ band. Finally, the
galaxy models were convolved with the observed PSF. The same procedure was used
to retrieve the structural parameters both in the simulated and actual images.

\begin{figure*} 
\centering
\plotone{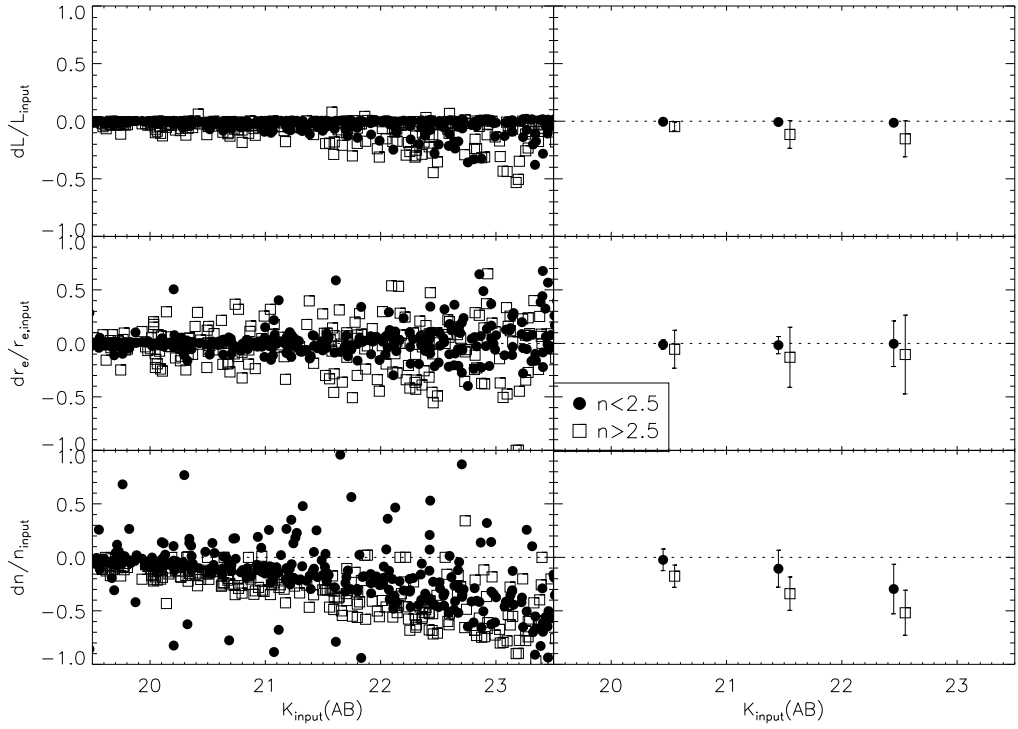}

 \caption{The relative error derived from the difference between the input and
recovered structural parameters ((output-input)/input) according to our
simulations for the FIRES MS1054 field. Solid symbols are used to  indicate
less concentrated objects (n$_{input}$$<$2.5) whereas open symbols imply highly
concentrated objects (n$_{input}$$>$2.5). The right column of plots shows the mean
systematic difference and 1 $\sigma$ error bars.} \label{renmag}

\end{figure*}

\begin{figure*}

\centering
\plotone{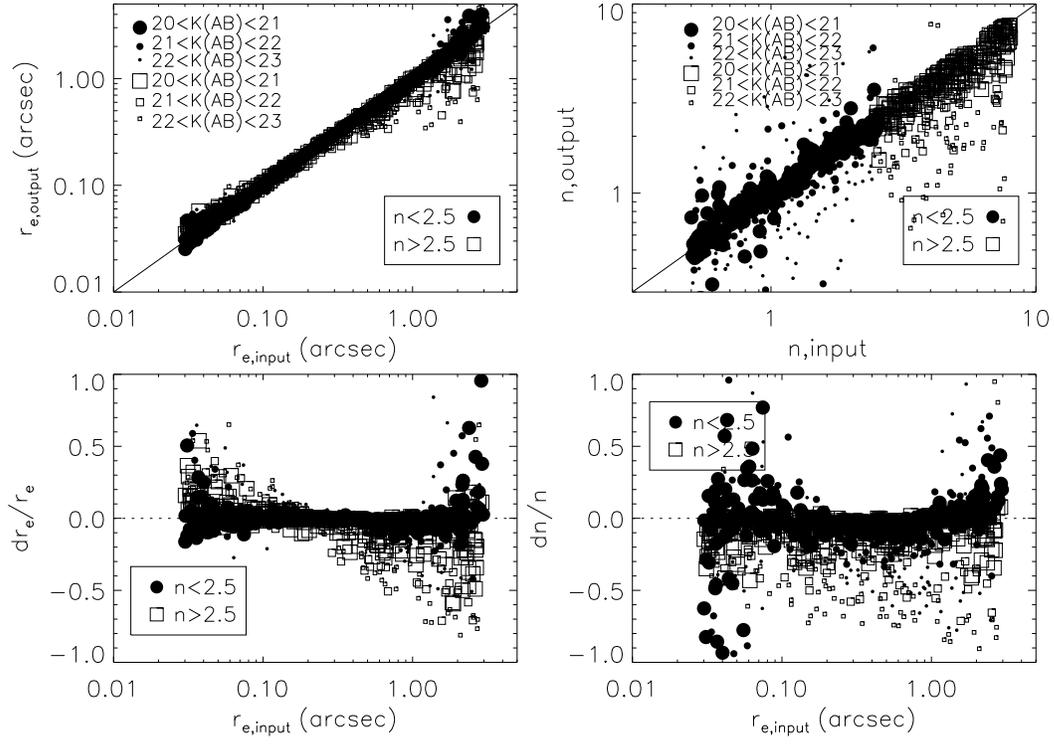}
\vspace{1cm}

  \caption{Galaxy size--measurement bias: The figure shows a comparison between
input and recovered structural parameter values in our simulations for the FIRES
observations of the MS1054 field. $Top$ $Left$: The relation between measured
and the input intrinsic half--light radius (before seeing convolution). $Top$
$Right$: The relation between measured and input seeing deconvolved S\'ersic
index $n$. $Bottom$ $Left$: The relative error between the input and the
measured seeing deconvolved effective radius
(dr$_e$/r$_e$=(r$_{e,output}$-r$_{e,input}$)/r$_{e,input}$) versus the input
effective radius. $Bottom$ $Right$:  The relative error between the input and
the measured seeing deconvolved S\'ersic index $n$
(dn/n=(n,output-n,input)/n,input) versus the input effective radius. Solid
symbols are used to indicate less concentrated objects (n$_{input}$$<$2.5) whereas
open symbols imply highly concentrated objects (n$_{input}$$>$2.5).}

\label{renversusren}

\end{figure*}

The results of these simulations are shown in Figs. \ref{renmag} and
\ref{renversusren}. Towards fainter apparent magnitude the parameters recovered
are systematically worse. At increasing magnitude the code recovers
systematically lower S\'ersic indexes. The bias  depends strongly on the shape
of the surface brightness profiles. We illustrate this by separating the
galaxies between less light concentrated profiles (n$_{input}$$<$2.5) and highly
concentrated profiles (n$_{input}$$>$2.5). Galaxies with larger $n$ are more
biased than those with lower values.


To illustrate the magnitude of the biases in the different parameters we
summarize the results for the most affected bin, K$_s$=22.5 mag.  For galaxies
with n$_{input}$$<$2.5 we find the following systematics: 1($\pm$3)\% lower
luminosities, 0($\pm$20)\% lower sizes, 30($\pm$23)\% lower S\'ersic indices.
For galaxies with n$_{input}$$>$2.5: 15($\pm$16)\% lower luminosities,
10($\pm$37)\% lower sizes, 52($\pm$21)\% lower S\'ersic indices. At brighter
magnitudes the structural parameters are recovered more accurately. 

As shown in Fig. \ref{renversusren} the systematic errors in the structural
fitting parameters depend on the apparent magnitude, r$_e$ and n. To facilitate
the discussion of these biases in our results (see Sect. \ref{robustness})  we
have quantified analytically what is the relation between the input and output
structural parameters depending on the magnitude, r$_e$ and n by fitting the
following expressions: r$_{e,out}$=$p_{re}$$\times$r$_{e,input}$$^{q_{re}}$ and
n$_{out}$=$p_{n}$$\times$n$_{input}$$^{q_{n}}$ to the results of our
simulations. The values of p and q obtained from the fittings are summarized in
Table \ref{corrections}. The difference between the input and output magnitudes
has also been quantified as a function of the input magnitude and the index n
(see first row of Fig. \ref{renmag}). It must be mentioned, however, that the
effect on the luminosities of our objects is very small $\lesssim$15\% in all
the cases (i.e.$\lesssim$0.15 mag). We correct the magnitudes according to the
following expression K$_{out}$=p$_m$+K$_{input}$.  The above expressions allow
us to transform from the three elements set (K$_{s,observed}$, r$_{e,observed}$,
n$_{observed}$) to (K$_{s,corrected}$, r$_{e,corrected}$, n$_{corrected}$). We
have used only the above corrections in Sect. \ref{robustness} to discuss how
robust are our results. The results shown in the rest of the paper are based on
the directly measured quantities without any attempt to correct the measured
parameters in order not to artificially increase the scatter.

It is important to note that although the seeing half--radius ($\sim$0.3\arcsec) is
similar to the effective radii of the galaxies we are dealing with, we can
estimate reasonable structural parameters due to the depth of our images. 
Galaxies at our K$_s$=23 mag analysis limit are a full 3 magnitudes brighter
than our 3$\sigma$ limit for point sources. This allows us to explore the
surface brightness radial profiles to 2.5--3 times the  seeing half--light
radius.

\subsubsection{Comparison between different filters}

Mock galaxies are useful to estimate the biases on the recovered
structural parameters. However, one can argue that because artificial
galaxies are simplistic representations of real galaxies, the errors
and bias determinations yield lower limits to the real case.  We have
checked the internal consistency of our data, comparing the size and
shape of our galaxies between the set of near infrared filters
used. The seeing and the depth are slightly different amongst the NIR
images which allows us to have a robustness test which is not based on
simulations. Naturally, this test is only useful under the assumption
that the change in the size and the shape of the light profile of the
galaxies due to changes in the wavelength along the set of NIR filters
is smaller than the intrinsic error in estimating the structural
parameters.

\begin{figure*}

\centering

\plotone{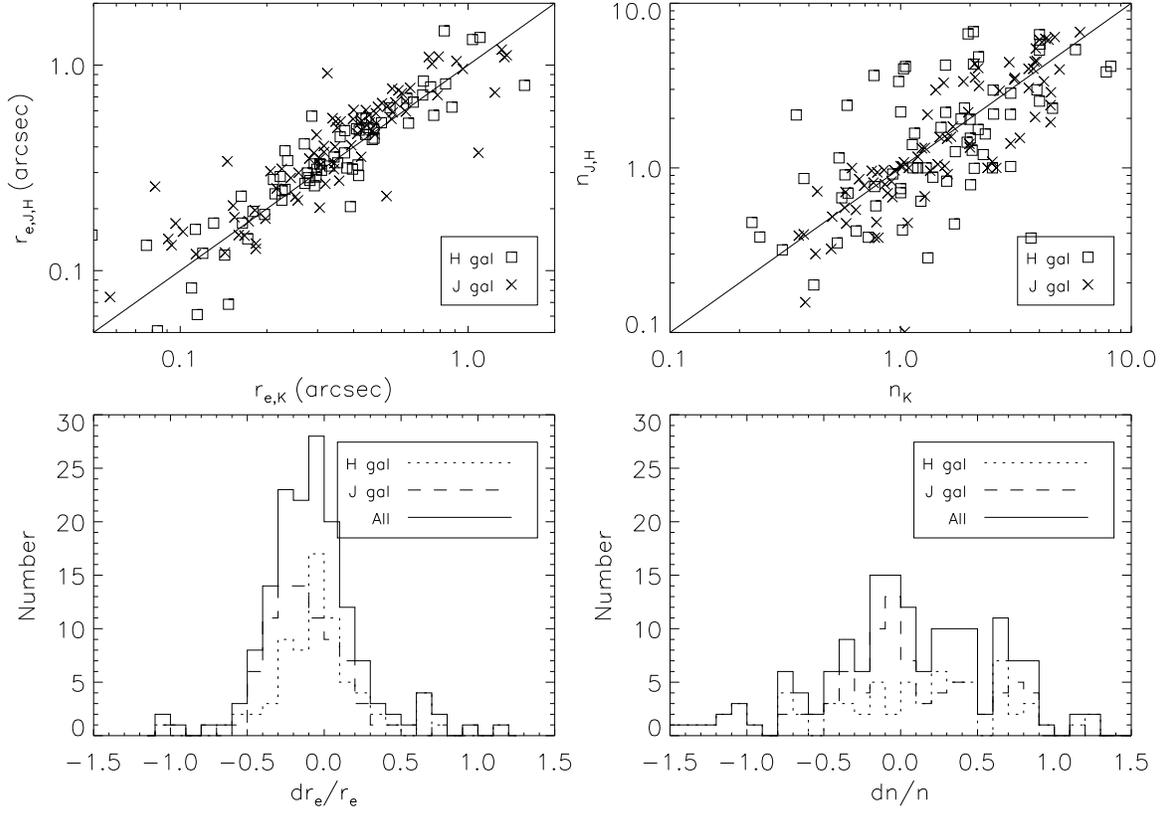}

\vspace{1cm}

  \caption{$Upper$ $panels$: Comparison between the profile shapes and size
estimates using the FIRES J$_s$ or H  filters versus the K$_s$ band for  all
the galaxies in the MS1054 field with 1$<$z$<$2.6. To match the rest--frame
optical  V--band, galaxies with 1$<$z$<$1.5 were observed in the J$_s$--band,
and galaxies with 1.5$<$z$<$2.6 were observed in the H--band. $Lower$ $panels$:
The relative difference between the size and the shape parameter measured in
the different filters:
dr$_e$/r$_e$=2$\times$(r$_{e,K}$-r$_{e,J,H}$)/(r$_{e,K}$+r$_{e,J,H}$) and
dn/n=2$\times$(n$_{K}$-n$_{J,H}$)/(n$_{K}$+n$_{J,H}$). The standard deviation
for the sizes is $\sim$24\% and for the shapes $\sim$60\%.}

\label{filtercomp}

\end{figure*}

Fig. \ref{filtercomp} shows the comparison between the sizes and the S\'ersic
indexes estimated in the K$_s$ band versus the sizes and the S\'ersic indexes
estimated using J$_s$ (1$<$z$<$1.5) and H (1.5$<$z$<$2.6) bands for galaxies of
the MS1054 field with 1$<$z$<$2.6. The sizes estimated using the different
filters present a $\sim$24\%  (1$\sigma$) of relative scatter between them
whereas the scatter for the shapes is larger ($\sim$60\%).

\subsubsection{Comparison using different PSFs}

\begin{figure*}

\centering

\plotone{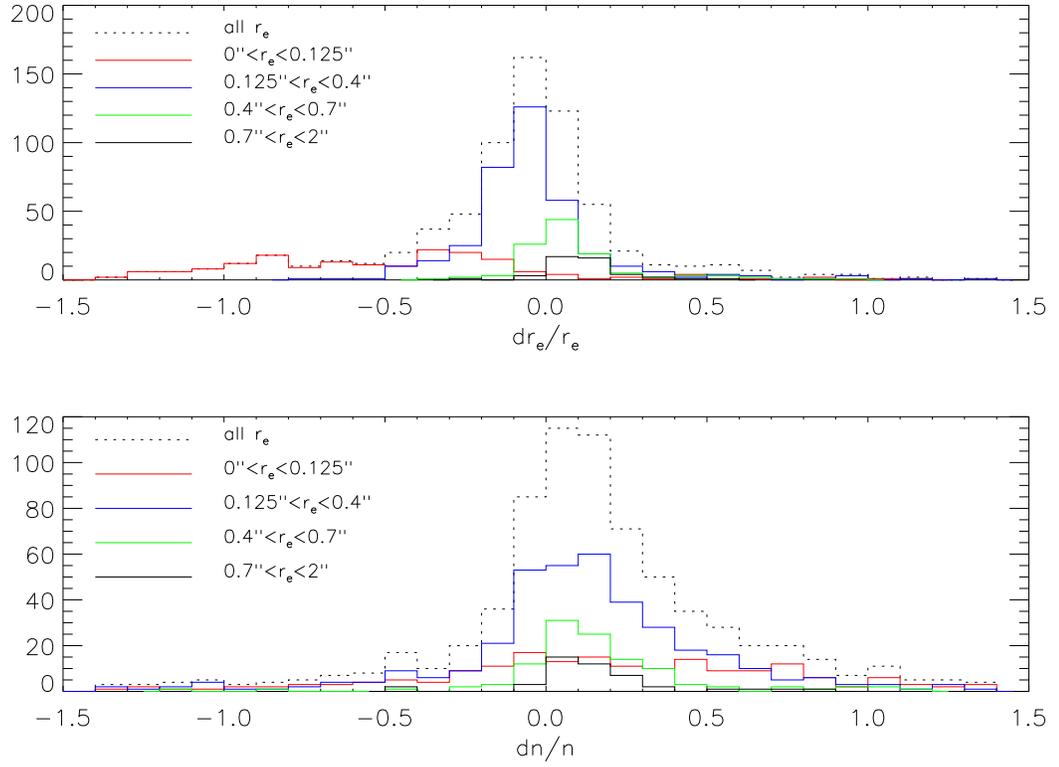}

  \caption{Reliability of the structural parameter estimation using different
PSFs. $Top$ $Panel$. The relative difference between the circularised sizes
estimated in the K$_s$ using  a PSF with a FWHM equal to the median value of the
different PSFs (PSF1) and the size measured using a PSF with a FWHM 2$\sigma$
times larger than the median (PSF2). 
dr$_e$/r$_e$=2$\times$(r$_{e,PSF2}$-r$_{e,PSF1}$)/(r$_{e,PSF2}$+r$_{e,PSF1}$).
$Bottom$ $Panel$. Same than in the top panel for the S\'ersic index n:
dn/n=2$\times$(n$_{PSF2}$-n$_{PSF1}$)/(n$_{PSF2}$+n$_{PSF1}$).}

\label{psftest}

\end{figure*}

We have explored also whether the variation of the PSF along the image can affect the
recovery of the structural parameters. To do that we have made a conservative
test reanalysing the full set of galaxies in the K$_s$ band using a PSF with a
FWHM 2$\sigma$ times larger than the value of the median FWHM of the PSFs. The
results of doing this are shown in Fig. \ref{psftest}. 

Only very compact galaxies with effective radii similar or smaller than the
pixel size are significantly affected by the change of the PSF along the field
of view. In those cases the estimation of the index n is pretty uncertain and we
can not allocate these galaxies to the low--n or high--n categories. These
objects amount to $\sim$20\% of our sample. According to their SEDs these
objects are not misidentified stars neither are they compatible with being at
z$<$1. Because of their extremely compact nature some of them could be AGNs. In
fact, for the  brightest object, MS1356, where spectroscopic analysis has been
made (van Dokkum et al. 2003;2004) the AGN hypothesis is confirmed. In that
case, their sizes could be not indicative of the sizes of their host galaxies.
However, we can not assure the AGN nature for all these objects, so we have
decided to explore how large could be  the effect of these objects in our
luminosity--size and stellar mass--size relations (see section 4.1 and 4.2). 

 For the rest of the sample ($\sim$80\% of our objects) the estimation
of the structural parameters is robust to changes in the selected PSF to analize
the data: the scatter between the sizes is $\lesssim$14\%(1$\sigma$) and the
scatter between the S\'ersic index n is $\lesssim$30\%(1$\sigma$).

\subsubsection{Size estimates at fixed n}

Another possible test to estimate the robustness of our  size estimations is to
reanalyze the objects using this time the S\'ersic index parameter fixed at $n$=1 or
$n$=4. We have repeated our analysis for the galaxies in the MS1054 field using the
filters which match the V--band rest--frame at every z. All the galaxies are fitted
initially with $n$ fixed to 1 and then refitted using $n$ equal to 4. From these two
fits we take that with the minimum $\chi^2$ value as representative of the galaxy
structural properties.

\begin{figure*}

\centering

\plotone{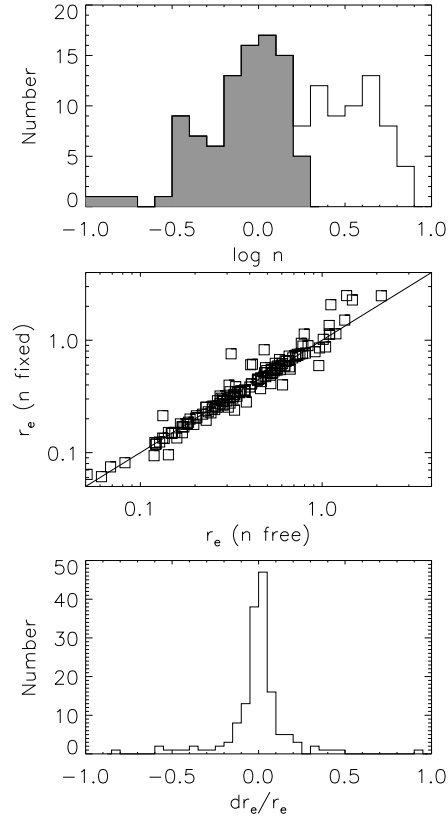}

\vspace{1cm}

\caption{$Top$ $Panel$.  The grey histogram shows the  S\'ersic index
distribution (when leaving this parameter free in the fitting process) for the
subset of galaxies which are better fit with a fixed S\'ersic parameter  to
$n$=1 whereas the open histogram shows the shape distribution for the galaxies
well fitted with $n$=4.  $Center$ $Panel$. The comparison between the size
estimated using $n$ free versus the size estimated using $n$ fixed to 1 or 4.
$Bottom$ $Panel.$ The relative difference between the size estimated using $n$
fixed or free: dr$_e$/r$_e$=2$\times$(r$_{e,n free}$-r$_{e,n fixed}$)/(r$_{e,n
free}$+r$_{e,n fixed}$). The scatter between both sizes estimates is $\sim$7\%
(1$\sigma$). The structural parameters
are estimated using the filters which match the V--band restframe at every z.}

\label{n1n4}

\end{figure*}

The comparison between the structural parameters recovered using $n$ fixed and $n$
free is shown in Fig. \ref{n1n4}. Galaxies better fitted by an exponential profile
($n$=1) have 0$<$n$<$2 when this parameter is left free during the fit. In addition,
galaxies well fitted by a de Vaucouleurs profile ($n$=4) yield $n$ ranging from 1.5
to 7. It is interesting to note that there is some overlap between both regimes
(1.5$<$n$<$2).  From the results presented here and in \ref{simparameters}, it seems
to be possible to discriminate between highly and less concentrated objects (i.e.
those with $n_{input}$ larger or smaller than 2.5 respectively) using
$n_{output}$=1.5 as the separation criterion. In fact,  if we assume, as suggested by
our simulations, up to a 50\% bias on the index $n$ for the high--concentrated
objects, an object with original $n_{input}$=3--4 would be identified in our code as
$n_{output}$=1.5--2. It is important to note that our criterion for separating the
galaxies using $n_{output}$=1.5 would be similar to using $n$=2.5 in a case where the
index $n$ was less biased than in the current analysis (see e.g. Barden et al.
(2005). In what follows, we will take advantage of this to facilitate a comparison of
our results with those found at lower z (see Sections 4 and 5.1).

The sizes estimated using $n$ fixed or $n$ free during the fit show very good
agreement with only $\sim$7\% (1$\sigma$) of relative scatter between them and
no significant bias.

\subsubsection{Comparison with NICMOS data}

We have  obtained deep H--band NICMOS images of the HDF--S. These NICMOS
data consist of 8 pointings of camera 3 (52"x52", 0.203"/pix).  Each pointing is the
combination of 6 sub-pixel dithered exposures, with a total exposure time of 1.5
hours.  The final mosaic was assembled using the {\it drizzle} task and has a pixel
scale of 0.119" to match our ISAAC ground-based data\footnote{The ISAAC pixel scale
is actually 0.''147; however, we resampled the ISAAC pixels to 3 $\times$ 3 blocked
HDF--S WFPC2 pixels.}. A detailed presentation of this dataset and an analysis of
the  sizes of the galaxies in this image will be presented in Zirm et al. (2006).

We have 27 galaxies in common between ISAAC and NICMOS images in the redshift
range 1.5$<$z$<$2.6 for which we analyze the H--band images. We found a good
correlation between the sizes measured in the NICMOS images compared with those
measured with ISAAC. The scatter is 24\%(1$\sigma$) with no systematic bias
between both measurements.

\subsection{Selection Effects}

In practice, any image presents a surface brightness limit beyond which the
sample is incomplete. To characterize this limit is particularly important for
high--z samples where the effects of the cosmological surface  brightness
dimming are severe. For a given total flux limit, the surface brightness limit
translates into an upper limit on the size for which a galaxy can be detected.

\begin{figure*}
\centering
\plotone{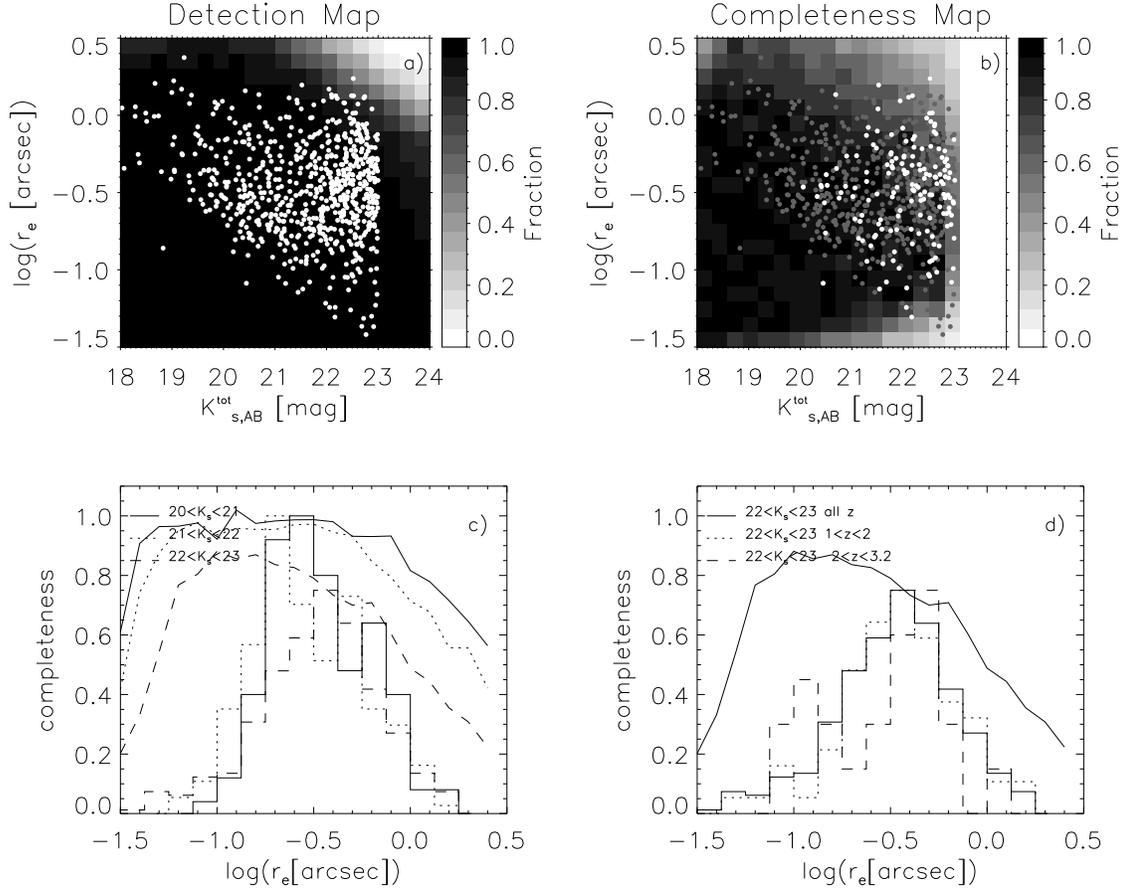}
  \caption{$a)$  Detection map for simulated sources with exponential profiles
placed at random in our K$_s$ band image of the MS1054 field. The grey-scale map
reflects the ratio between input and recovered objects per input magnitude  and
$\log$(r$_e$) bin. Overplotted on the map is the distribution of the full sample
of K$_s$ band selected objects in the MS1054 field.  $b)$ Completeness map for
simulated sources with exponential profiles  placed at random in our K$_s$ band
image of the MS1054 field. The grey-scale map reflects the ratio between the
number of output galaxies  with recovered magnitude and size at a given
magnitude and  $\log$(r$_e$) bin and the number of input galaxies with input
magnitude and size in that bin. Overplotted on the map is the distribution of
the full sample of K$_s$ band selected objects in the MS1054 field with those
explored in this paper (1$<$z$<$3.2) highlighted. $c)$ The completeness for
three different magnitude intervals: 20$<$K$_s$$<$21, 21$<$K$_s$$<$22 and
22$<$K$_s$$<$23 as a function of the size (smooth curves). Overplotted are the
size distributions (arbitrarily normalized to have a value at the peaks equal to
the completeness value provided by the completeness curve at that r$_e$) of real
galaxies in the same intervals (histograms).  $d)$ The completeness for our
faintest magnitude interval   22$<$K$_s$$<$23 as a function of the size (smooth
curve). Overplotted are  the apparent size distributions (arbitrarily normalized
to have a value of 0.75 in the peak) of  real galaxies in the same interval
(histograms) for: all the galaxies, galaxies with 1$<$z$<$2 and galaxies with
2$<$z$<$3.2. The apparent size distribution of the galaxies in this magnitude
interval is independent of redshift. The observed size distribution decline more
rapidly to larger sizes than the completeness limit. This indicates that our
sample is not significantly affected by incompleteness of the largest galaxies
at a given magnitude.}
\label{completeness}
\end{figure*}

To determine the detection map of the FIRES MS1054 K$_s$--band image we have
created a set of 10$^5$ mock sources with intrinsic exponential profiles
uniformly distributed as follows: K$_s$--band total  magnitudes between 18 and
24 mag, effective radius r$_e$ between 0.03 and 3 arcsec  and inclination angles
between 0 and 90 degrees. Readers more interested in the simulations are
referred to F\"orster Schreiber et al. (2006)\footnote{Simulations shown in
F\"orster Schreiber et al. (2006) only consider point sources with an input
magnitude distribution following the slope of the counts.}. The simulated
sources are placed randomly on the real image 20 at a time and extracted as for
the real source detection. On doing that we construct a detection map giving the
number of recovered sources over the number of input artificial sources per
input magnitude and input $\log$(r$_e$) bin (see Fig. \ref{completeness}a). A
equivalent analysis for the HDF--S field is presented in Fig. 8 of Trujillo et
al. (2004).  It is important to note that in selecting exponential profiles
($n$=1) for estimating our detection map we are being conservative from a
detection standpoint. Galaxies with larger $n$, and consequently more centrally
concentrated, would be much easier to detect at a given magnitude.

We have also estimated the completeness map (see Fig. \ref{completeness}b) of
our survey for those galaxies with measured magnitude K$_s$$<$23. To do that we
have computed the ratio between the number of recovered sources with  output
magnitude and output size over the number of input sources within that magnitude
and size bin. To estimate the output magnitudes and sizes we have used exactly
the same tools as for actual galaxies. Overplotted on the completeness map is
the distribution of the full sample of K$_s$ band selected objects in the MS1054
field. Highlighted in this distribution are those objects which are used in this
paper (i.e. those with 1$<$z$<$3.2).ste

As a second step to analyze the effect of completeness in our sample we have
probed whether the size distribution of our objects could be affected by the
completeness. In Fig. \ref{completeness}c we show the completeness for three
different magnitude intervals: 20$<$K$_s$$<$21, 21$<$K$_s$$<$22 and
22$<$K$_s$$<$23 as a function of the size. In addition, we overplot the size
distribution (arbitrarily normalized to have a value at the peak equal to the
value of the completeness curve at that point) of real galaxies in the same
intervals. The number of observed galaxies decreases more rapidly to larger
sizes than do the completeness curves. This shows that incompleteness is not
affecting the extent of our size distribution to larger sizes. A similar
analysis but this time using only the faintest magnitude bin (22$<$K$_s$$<$23)
is done separating the galaxies according to their redshift (Fig.
\ref{completeness}d). This figure shows that the size distribution of the
observed galaxies in the magnitude interval 22$<$K$_{s,input}$$<$23 is not
related with the redshift of the objects. Interestingly,  Bouwens et al. (2004)
show, using UDF images, that the principal effect of increased depth is to add
galaxies at fainter magnitudes, not larger sizes, demonstrating that high--z
galaxies are predominantly compact and that large low surface brightness objects
are rare. This result provides independent corroboration of our analysis. The
effect of the completeness in the robustness of our relations is explored in
Sec. \ref{robustness}.

The interested reader could also see how the size distribution of the SDSS
galaxies would look like under the FIRES sample selection effects (Trujillo et
al. 2004; their section 4.1). The depth of our images ensures that the largest
SDSS galaxies would be detected if they were present in our sample.

We have also quantified the mass and luminosity limits implied by our observed
magnitude limit.  In doing so we try to serve the dual purpose of maximizing the
number of objects in our sample while simultaneously reducing systematic biases on
the final results.  We determine our rest-frame luminosity limit using the K$_s$
magnitude and the expected color of an Scd template at $z=2.5$, the center of our
highest redshift bin.  For K$_s=23.5$ this limit is ${\rm L_V}>3.4\times 10^{10} {\rm
h_{70}^{-2} L_{\odot}}$.  Above this limit we are complete at all redshifts
$z\lesssim2.5$ in the HDF--S field.  We adopt the same limit for the MS1054 data
acknowledging that we will be missing galaxies in our higher redshift bin with
$23<{\rm K}_s<23.5$. As shown in Figs. \ref{lumreall} and \ref{massreall}, however,
the distributions in size, luminosity, and mass of objects in the MS1054 and the
HDF--S fields are similar and we make the assumption that this incompleteness in the
highest redshift MS1054 data will not significantly bias our results.

\begin{figure*}

\centering

\plotone{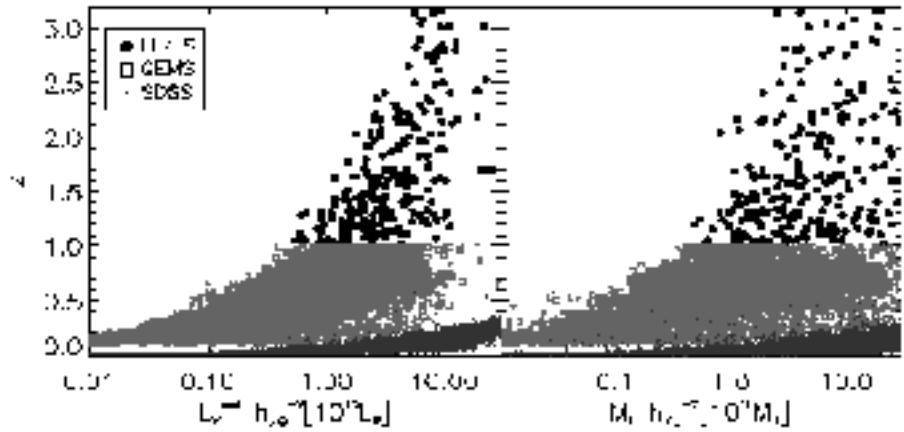}

\caption{The L$_V$--z and M$_*$--z diagrams for the combined data set used in
the present analysis. Solid points correspond to the FIRES galaxies in the
HDF--S and the MS1054 fields, open squares are GEMS galaxies  (McIntosh et al.
2005; Barden et al. 2005) and dots are the SDSS galaxies (Shen et al. 2003).
Only the most luminous and the most massive objects can be homogeneously
explored along the full redshift range. Since the mean  redshift is our highest
redshift bin is $\sim$2.5, only galaxies with 
L$_V$$\gtrsim$3.4$\times$10$^{10}$h$_{70}$$^{-2}$L$_\odot$  can be studied as a
homogeneous sample. Objects with the lowest mass--to--light ratios can be
homogeneously explored if their masses are
M$_\star$$\gtrsim$3$\times$10$^{10}$h$_{70}$$^{-2}$M$_\odot$. We are complete
to objects of every stellar mass--to--light ratio if
M$_\star$$\gtrsim$6.6$\times$10$^{10}$h$_{70}$$^{-2}$M$_\odot$ (see text for
details).}

\label{lumzmassz}

\end{figure*}

We choose two separate means of defining a limit in mass.  For our first mass
limit we choose the lowest observed mass in our combined sample at $z\sim 2.5$
(see Fig. \ref{lumzmassz}).  This limit is ${\rm M_*>3\times 10^{10} {\rm
h_{70}^{-2} M_{\odot}}}$.  We realize that only the objects with the lowest
mass-to-light ratios will be detectable at these masses and that we are
incomplete to objects of higher mass-to-light ratios. Nonetheless we use this
limit to maximize the total number of objects in our sample, keeping in mind
that we may experience systematic biases from our mass incompleteness.  As a
more conservative approach we also choose a mass limit corresponding to the
maximum stellar mass-to-light ratio expected at $z\sim 2.5$.  We use a
maximally old single stellar population from Bruzual \& Charlot (2003) with
solar metalicity and a Salpeter (1955) IMF.  At $z\sim2.5$ the Universe is
$\sim 2.6$ Gyr old for our cosmology and the resultant mass-to-light ratio is
1.93.  Coupled with our luminosity limit of $3.4\times 10^{10} {\rm h_{70}^{-2}
L_{\odot}}$, this yields a mass limit of ${\rm M_*>6.6\times10^{10} h_{70}^{-2}
M_{\odot}}$.  Above this limit we are complete to objects of every stellar
mass--to--light ratio, although we have very few objects and our random errors
will be large.  The differences between results using these two limits are
discussed in the end of \S~6.  As done for the luminosity threshold we adopt
the limits for the HDF-S for the whole sample.

\section{THE OBSERVED LUMINOSITY/STELLAR MASS $VS$ SIZE RELATIONS AT HIGH-Z}

\subsection{Luminosity vs size}

We now present the relation between  luminosity  and the rest--frame V--band
size, covering the redshift range 1$<$z$<$3.2 for  the HDF--S and the MS1054
fields. The low redshift limit is selected to avoid the influence of cluster
galaxies at z=0.83 in the MS1054 field and the high redshift limit is chosen to
maintain our analysis of the high--z galaxies in the optical rest--frame.   We
convert our measured angular sizes to physical sizes using the photometric
redshift (or the spectroscopic value when available) determined for each object.

\begin{figure*}

\centering

\plotone{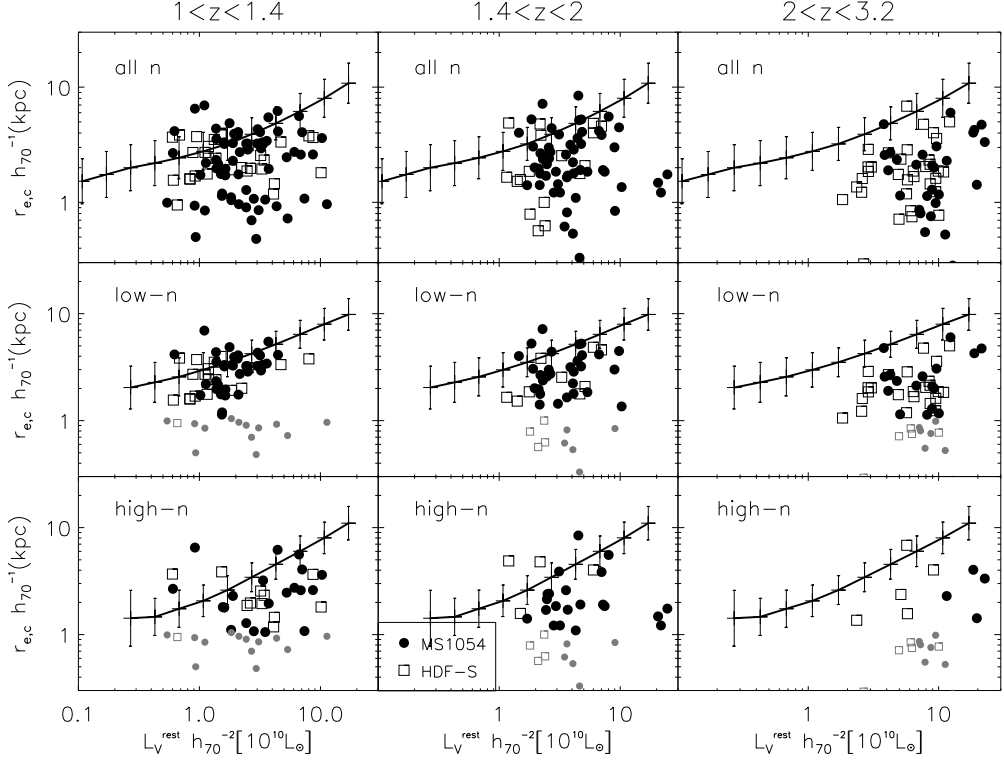}

  \caption{Distribution of the rest--frame optical sizes vs. the  rest--frame
V--band luminosities for all galaxies from FIRES. Galaxies from the HDF--S
field (Labb\'e et al. 2003) are shown by open squares and galaxies from the
MS1054 field (F\"orster Schreiber et al. 2005) by filled circles. The different
rows show the galaxies separated according to their S\'ersic index
concentration parameter. For objects with r$_e$$<$0.\arcsec125 the
estimation of the S\'ersic index n is uncertain. For that reason,
these objects are plotted simultaneously in the low and high-n rows using
lighted symbols. Overplotted on the observed distribution of points are
the mean and dispersion of the distribution of the S\'ersic half--light radius
of the SDSS galaxies (in the ``V--band'') as a function of the V--band
luminosity. The second and third row show the SDSS distributions separating
into late and early type respectively. For clarity individual error bars for
the FIRES data are not shown; the mean size relative error is 25\%.}

\label{lumreall}

\end{figure*}

In Fig. \ref{lumreall} our sample is split in three different redshift
bins: 1$<$z$<$1.4, 1.4$<$z$<$2 and 2$<$z$<$3.2. This separation allows
us to study the galaxies in roughly equal time intervals of $\sim$1.2
Gyr. 

The top row shows the luminosity--size relation for the full sample. The middle
row and the bottom row show the same relation but this time separating the
galaxies by their  concentration. For objects with r$_e$$<$0.\arcsec125 the
estimation of the S\'ersic index n is uncertain. To indicate  this incertitude
these objects are plotted simultaneously in the low and high-n rows using
lighted symbols.

  Overplotted on our observed distributions are the mean and dispersion of the
distribution of the S\'ersic half--light radii from the Sloan Digital Sky
Survey (SDSS; York et al. 2000) galaxies. We use the ``local'' SDSS sample  for
reference. The sizes are determined from a S\'ersic model fit (Blanton et al.
2003). The characteristics of the sample used here are detailed in Shen et al.
(2003). The mean of the SDSS galaxies redshift distribution  used for
comparison is 0.1. We use the sizes and the shapes estimated in the observed
r--band as this closely matches the V--band restframe filter at z$\sim$0.1. The
luminosity of the SDSS galaxies in the restframe V--band are estimated by
interpolating between the restframe g--band and  r--band luminosities (S. Shen,
private communication).

In the first row, our sample is compared to the total population observed by SDSS,
whereas in the second row we compare with the galaxies classified by Shen et al. as
late--type and in the third row with those classified as early--type. Their early or
late--type classification is based on the S\'ersic index: galaxies with n$<$2.5 are
considered late--types and galaxies with n$>$2.5 are identified as early--types. It
is important to note that using even smaller index n values like n=2 as the criterion
for the separation between early-- and late--type galaxies in the SDSS does not
produce a significant change in the luminosity-- and  stellar mass--size relations
(S. Shen, private  communication). This is as expected because of the scatter between
the S\'ersic index $n$ and the Hubble Type relation (see e.g. Fig. 1 of Ravindranath
et al. 2004). Consequently, changing from n=2 to n=2.5 (or vice versa) does not
change substantially the morphological type of the galaxies under study, and
therefore, the effect on the luminosity--size or stellar mass--size relations is
small.

Returning now to the redshift evolution, Fig. \ref{lumreall} shows that at a given
luminosity, galaxies are progressively smaller at higher z. Of course, this
evolution of the luminosity--size relation can be interpreted differently: at a
given size, galaxies were more luminous at higher z.

\begin{figure*}

\centering

\plotone{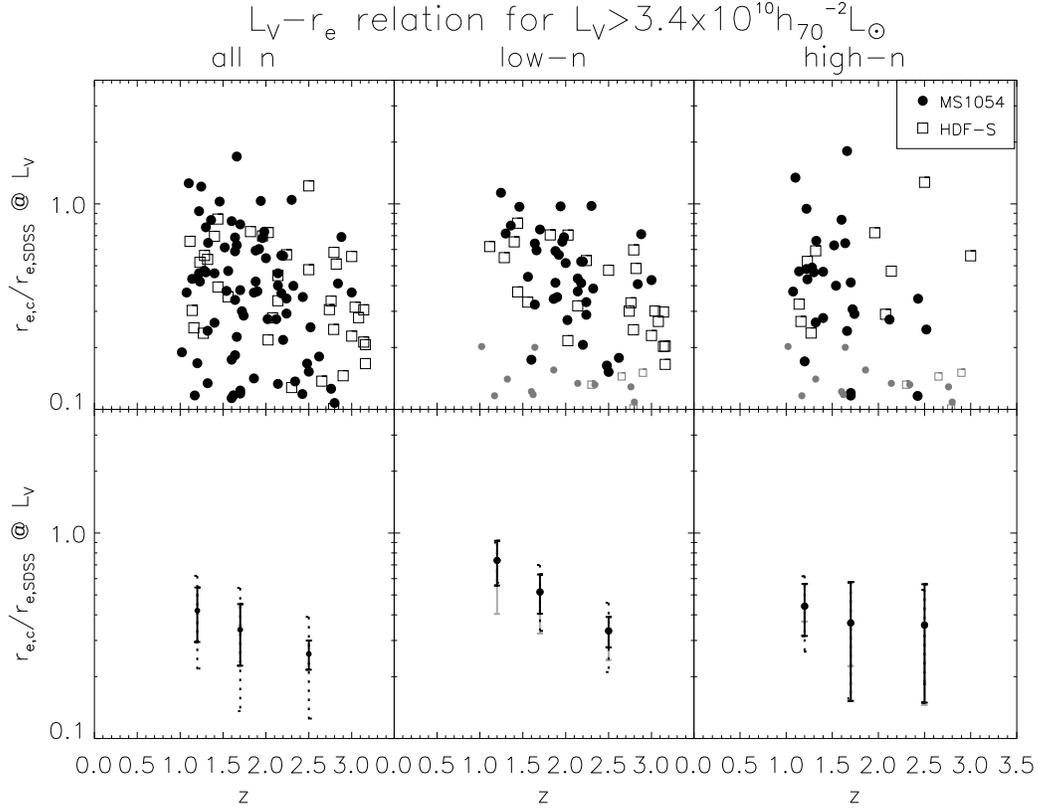}

  \caption{Redshift evolution of the size--luminosity relation for FIRES
galaxies: the figure shows the ratio between the observed size  (at a given
luminosity) and the mean size of equally luminous present--day galaxies from the
local SDSS sample as a function of z. For objects with r$_e$$<$0.\arcsec125 the
estimation of the S\'ersic index n is uncertain. For that reason, these objects
are plotted simultaneously in the low and high-n rows using lighted symbols. The
upper panels show the individual objects whereas the lower panels show the
dispersion (dotted error bars) and the uncertainty (2  $\sigma$) in the mean
determination (solid error bars) estimated from the
$\log$(r$_{e,c}$/r$_{e,SDSS}$) distribution. Grey error bars show how the
contribution of the small galaxies could affect the estimation of the mean.  The
figure shows that galaxies of a given luminosity were physically smaller at
early epochs (or higher redshift). Alternatively, the plot shows that galaxies
of a given size were more luminous at higher z.}

\label{lumreallnzeta}

\end{figure*}

\begin{figure*}

\centering

\plotone{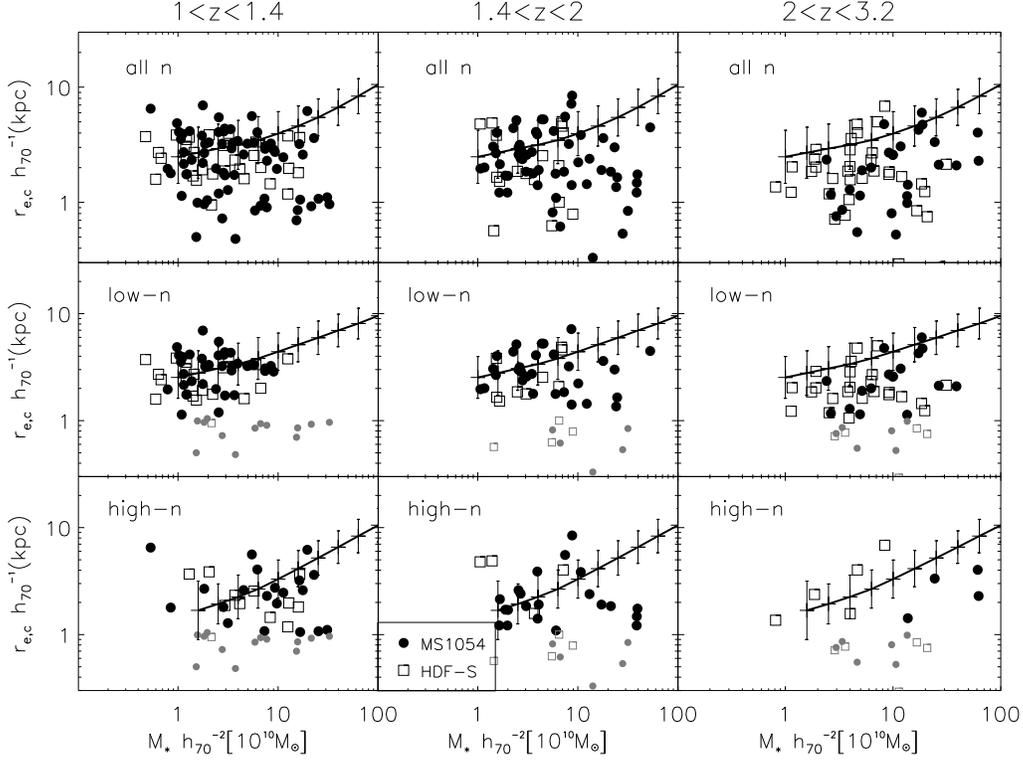}

  \caption{Distribution of rest--frame optical sizes vs. the stellar masses for
FIRES galaxies. Analogously to Fig. \ref{lumreall} galaxies from the HDF--S
field are shown by open squares and galaxies from the MS1054 field by filled
circles. The different rows show the galaxies separated according to their
S\'ersic index shape parameter. For objects with r$_e$$<$0.\arcsec125 the
estimation of the S\'ersic index n is uncertain. For that reason,
these objects are plotted simultaneously in the low and high-n rows using
lighted symbols. Overplotted on the observed distribution of
points are the mean and dispersion of the distribution of the S\'ersic
half--light radius of the SDSS galaxies  as a function of the stellar mass. The
second and third row show the SDSS distributions separated into late and early
type respectively. For clarity, individual error bars are not shown. The mean
size relative error is 25\%.}

\label{massreall}

\end{figure*}

To quantify the  evolution of these relations as a function of redshift, we show
in Fig. \ref{lumreallnzeta} the ratio between the observed size and the expected
size (at a given luminosity) from the SDSS distribution versus z. To estimate
the expected size from SDSS at a given luminosity  we interpolate linearly
between the SDSS points when necessary. From this plot the evolution in size (at
a given luminosity) with z is evident. Galaxies with
L$_V$$\gtrsim$3.4$\times$10$^{10}$h$_{70}$$^{-2}$L$_\odot$  at z$\sim$2.5 are
$\sim$3.5 times smaller than for equally luminous galaxies today.  In the second
row of this figure we show the evolution of the mean and the dispersion (large
error bars) of the above ratio estimated from the $\ln$(r$_{e,c}$/r$_{e,SDSS}$)
distribution. These quantities are estimated in  the same redshift bins as
stated above. The small error bars enclose the 2 $\sigma$ uncertainty of the
means. To evaluate these error bars we have used a bootstrapping method.

As in Fig. \ref{lumreall}, those galaxies with r$_e$$<$0.\arcsec125 are plotted
with lighted symbols. To measure how much these small galaxies could affect  the
luminosity--size evolution we have made the most conservative approach we can
do. First, we have assumed  that all those galaxies are in the low-n bin and we
have reestimated the mean value of the $\log$(r$_{e,c}$/r$_{e,SDSS}$)
distribution accounting for the contribution of the small galaxies. The range of
variation of the mean is shown with the grey error bar. In a second step, we
have assumed that all those galaxies belong to the high-n bin and we have
repeated the same exercise.

\subsection{Stellar mass vs size}

\begin{figure*}

\centering

\plotone{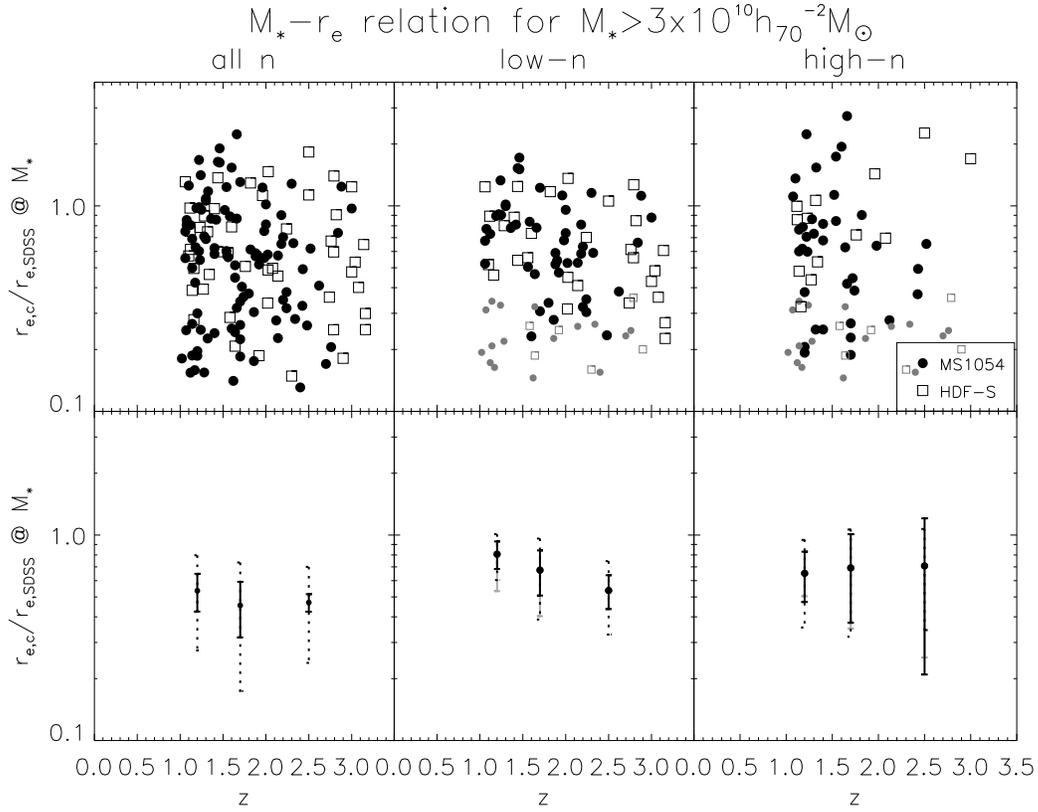}

\caption{The ratio between observed size of FIRES galaxies and the size (at a
given stellar mass) expected from the local SDSS sample shown as a function of
z. The upper panels show the individual objects whereas the lower panels show
the dispersion (dotted error bars) and the uncertainty (2 $\sigma$) at the mean
determination (solid error bars) estimated from the
$\log$(r$_{e,c}$/r$_{e,SDSS}$) distribution. Grey error bars show how the
contribution of the small galaxies could affect the estimation of the mean. The 
size at a given mass  evolves moderately with z.}

\label{massreallnzeta}

\end{figure*}

We have also explored the relation between stellar mass and size for our sample
(Fig. \ref{massreall}).  The stellar mass--size distribution evolves less than
the luminosity--size relation at high--z. The stellar mass--size relation
presents more scatter than the luminosity--size relation because the stellar
mass is an indirectly inferred property. This scatter  is ultimately related to
the uncertainty in the M/L determinations for these galaxies.  The evolution
with redshift of the sizes of the galaxies at given stellar mass is illustrated
in Fig. \ref{massreallnzeta} where we show the ratio between the observed size
and the expected size (at a given stellar mass) according to the SDSS local
sample. The potential contribution of the small galaxies to this relation is
estimated as for the luminosity--size relation.

The SDSS stellar masses used in Shen et al. (2003) are derived from stellar
absorption line indices centered on the inner region of the galaxies whereas
the present work uses colors integrated over the full galaxy. As discussed in
Kauffmann et al. (2003) this difference in techniques is particularly important
for brighter galaxies as they have strong color gradients, such that the
central colors are not indicative of the luminosity weighted total colors.
According to that work the mass--to--light ratio derived from line indices are
biased to higher values than those measured from integrated colors. To avoid
this problem, we have re--estimated the stellar masses of SDSS for this work
using the restframe (g--r) color  (S. Shen, private communication) and applying
the transformation suggested for this color in Bell et al. (2003). This
transformation is based on a Kroupa (2001) IMF. To match their values with the
FIRES data (which uses a Salpeter IMF) we apply the transformation suggested in
Kauffmann et al. (2003): M$_{IMF, Salpeter}$= 2$\times$M$_{IMF, Kroupa}$.

\subsection{Robustness of the Luminosity--size and stellar mass--size estimates}
\label{robustness}

\begin{figure*}

\centering

\plotone{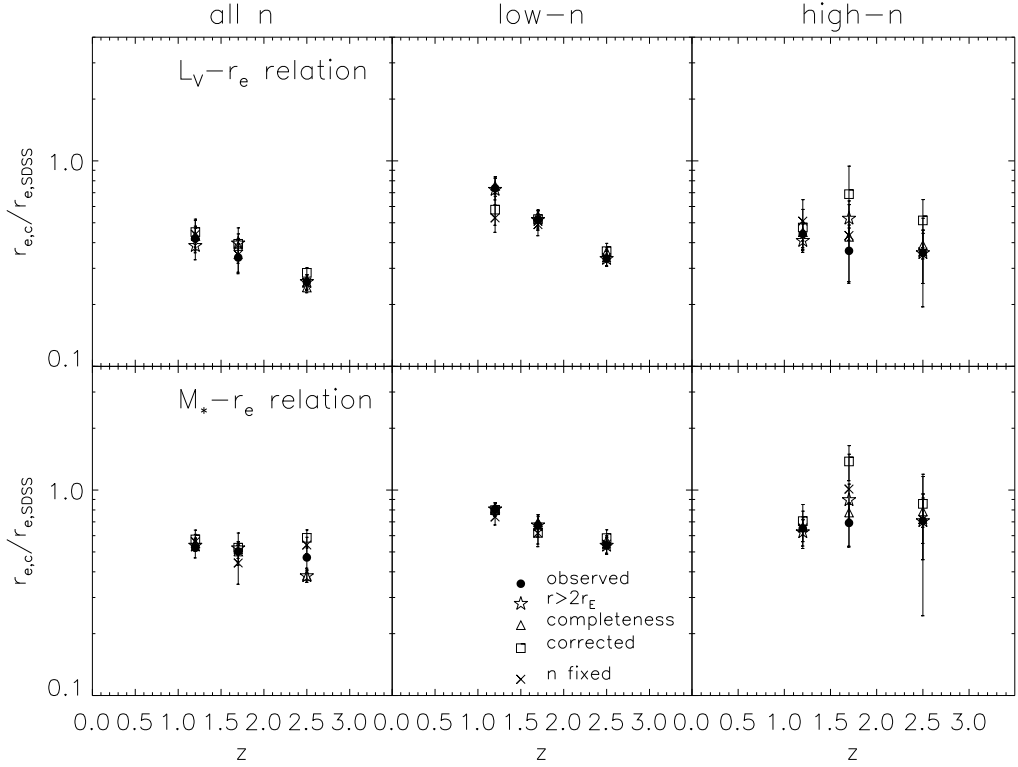}

  \caption{Comparison between five different  estimates of the mean luminosity-- and
  stellar mass--size distributions: the direct  estimates (solid points), the
  estimates omitting  the galaxies inside two Einstein radii (r$_E$$\sim$15'';
  Hoekstra et al. 2000) of the MS1054 cluster (open stars),  the estimation 
  weighting every galaxy according to the completeness map (open triangles), the
  estimation using the corrections suggested from our simulations (open squares) and
  the estimation using fits where the S\'ersic index $n$ is fixed to 1 or 4
  (crosses). The error bars show the 1 $\sigma$ uncertainty in estimating the mean of
  the distributions. All the points are in agreement within $\sim$1 $\sigma$. For
  clarity, bars showing the intrinsic dispersion of the relations are not included.}

\label{centeraccuracy}

\end{figure*}

The luminosity-- and stellar mass--size relations presented in the previous sections are
based on our direct measurements without making any attempt to correct for
possible biases in the structural parameters as indicated by the simulations. To
check whether the presented results are robust we have repeated our analysis
correcting this time the observed structural parameters following the
indications of our simulations (Table \ref{corrections}). In this particular
case, the separation between low--n and high--n galaxies is done
using n=2.5 as the separation criterion. In addition, we have also repeated our
analysis using the size estimation from the fits using $n$ fixed. We summarize
the results of these tests on Fig. \ref{centeraccuracy}.

As expected, due to the smaller sub--sample of galaxies  and the larger
corrections suggested by the simulations, the least robust results are for
galaxies with the larger light concentration (high--n). However, it is
interesting to note that all the estimates  of the mean relation are in
agreement within $\sim$1 $\sigma$. As most of our galaxies have a small index n
value, the corrections are small for most of the sample. Consequently, the
relations using the corrections suggested by the simulations do not change our
main results. In addition, when we compare our relations using $n$ free with
those obtained using $n$ fixed to $n$=1 or $n$=4, we  do not observe systematic
effects.

We have also studied whether the weak magnification lensing of the MS1054--03
foreground cluster can affect the result of our analysis. The cluster mass
distribution has been modeled by Hoekstra, Franx \& Kuijken (2000). The average
background magnification effects over the field of view covered by FIRES
observation range from a few \% to 25\% between z=1 to 4. The magnification is
most significant  in the immediate vicinity of the cluster central region. The
Einstein radius r$_E$ of this cluster is estimated to be $\sim$15 arcsec. We
have removed from our sample all the galaxies located within 2r$_E$ (this
implies 9 objects). Outside this region the magnification is expected to be
very  small. The result of removing these galaxies in our relations is shown in
Fig. \ref{centeraccuracy}. As expected from the small number of objects within
2r$_E$ the effect on our relations is very tiny. 

Finally, we have explored the effect of the completeness in our relations. To do
this we have weighted every galaxy of our sample with the inverse value provided
by our completeness map at every magnitude and size bin. The relations obtained
using the weights are shown in Fig. \ref{centeraccuracy}. As that figure shows,
due to the high completeness of our sample, the observed relations remain
basically unchanged. It should be noted, however, that our completeness map is
strictly valid only under the assumption of an uniform input distribution with
all the galaxies  well described by an exponential profile. This assumption
is realistic for $\sim$65\% of our sample.

The above tests indicate that the results presented in this paper are robust. 
Because the main results of this paper are insensitive to the corrections, we
perform our analysis based purely on the direct measurements. Applying these
corrections artificially increases the scatter of our relations because of the
necessary approximations when correcting. We find that the increase of the
scatter is $\sim$20--40\% in the corrected distributions related to those based
on the direct estimations.

\subsection{Robustness of the local SDSS relations}

Our analysis of the evolution of the luminosity--size and stellar mass--size
relations with redshift depends on the accuracy of the Shen et al. (2003) SDSS
local relations. Driver et al. (2005) have pointed out, using the
Millennium Galaxy Catalog (MGC), that surface brightness selection could bias
the Shen et al. results. Driver et al. (their Fig. 19) show an uniform offset
of $\delta\mu^e$$\sim$0.4 mag arcsec$^{-2}$ in the luminosity--surface
brightness distributions between their estimations and the Shen et al.
relations. At a given luminosity, the global distribution of galaxies in the
Shen et al. data presents a mean surface brightness $\sim$0.4 mag arcsec$^{-2}$
brighter than in the Driver et al. work. If we translate this into effective
radii this would imply that Shen et al. mean effective radius estimations are
(at a given luminosity) a factor 10$^{-0.2\delta\mu^e}$ (i.e. $\sim$0.83)
smaller  than the Driver et al. values. To account (crudely) for this offset in
our size evolution estimations we would need to multiply the values presented
in Table \ref{sizeevoldata} by the above factor. In this sense, the evolution
reported in this paper would be slightly less strong ($<$20\%) than the
evolution estimated using the MGC data as a reference. In any case, it is worth
noting that the main results of our papers would be basically unchanged by this
potential offset.

Similarly, we have also estimated the mean offset in size at a given luminosity
between the very low redshift (z$<$0.05) SDSS sample from Blanton et al. (2005)
and the Shen et al. relations. We have done this for brightest population
(L$_V$$\gtrsim$3.4$\times$10$^{10}$h$_{70}$$^{-2}$L$_\odot$). For these galaxies
we found $<$re$_{Shen}$/re$_{Blanton}$$>$=0.86. This value is similar to that
reported above comparing with the MGC galaxies, however, in this case the
difference must be taken with caution as it could be slightly affected by
potential evolution of the mean size of the galaxies since z$\sim$0.1 (Shen et
al.) to z$\sim$0 (Blanton et al.).

\section{ANALYSIS}

\subsection{Comparison of FIRES data to the evolution at z$<$1}

Several analyses of the luminosity--size evolution of galaxies in the optical
rest--frame up to z$\sim$1 have been carried out (Im et al. 1996; 2002, Lilly
et al. 1998, Schade et al. 1999, Simard et al. 1999, Ravidranath et al. 2004;
Trujillo \& Aguerri 2004; McIntosh et al. 2005; Barden et al. 2005). These
studies seem to agree on a moderate decrease of the surface brightness of the
galaxies towards the present: $<$1 mag in the V--band restframe (or
equivalently an increase in size at a given luminosity  of $\lesssim$35\%).

In order to make a consistent comparison at lower redshifts with  FIRES,  we
use the data from  the largest sample currently available at
intermediate redshift: the GEMS survey (Rix et al. 2004). GEMS is a large-area
(800 arcmin$^2$) two--color (F606W and F850LP) imaging survey with the ACS on
the HST  to a depth of m$_{AB}$(F606W) = 28.3(5$\sigma$) and m$_{AB}$(F850LP) =
27.1(5$\sigma$) for compact sources. Focusing on the redshift range
0.2$\leq$z$\leq$1, GEMS provides morphologies and structural parameters for
nearly 10,000 galaxies for which  redshift estimates, luminosities, and SEDs
exist from COMBO-17 (Classifying Objects by Medium--Band Observations in  17
Filters; Wolf et al 2001, 2003).

The luminosity--size and stellar mass--size relations of this survey are
presented in Barden et al. (2005; late--type galaxies) and McIntosh et al.
(2005; early--type galaxies). The GEMS late-- and early--type separation
criteria is based on the S\'ersic index $n$.  Late--types are defined through
n$<$2.5, and  early--types through n$>$2.5 and  a color within the
``red--sequence'' (Bell et al. 2004). We have checked that adopting smaller
index n values like n=2 instead of n=2.5 as the separation criterion does not
produce a significant change in their results.   The stellar masses of the GEMS
survey used in the present work are derived in the same way as those in
FIRES\footnote{In Barden et al. (2005) and McIntosh et al. (2005) the GEMS
stellar masses are also estimated from stellar populations models, finding no
differences in the resulting stellar mass--size relation.}. Using their
measurements of size, luminosity, mass, redshift and completeness we have
repeated the same analysis as for the FIRES sample. To ensure homogeneity with
the FIRES sample we have only selected GEMS galaxies with
L$_V$$>$3.4$\times$10$^{10}$h$_{70}$$^{-2}$L$_\odot$ (in the case of the
luminosity--size relation) and
M$_\star$$\gtrsim$3$\times$10$^{10}$h$_{70}$$^{-2}$M$_\odot$ (in the case of the
stellar mass--size relation). The resulting size evolution from both surveys
together are shown in Fig. \ref{gemsfires} and Table \ref{sizeevoldata}.

 From this comparison we see that the z$<$1 evolution (GEMS) and z$>$1
evolution (FIRES)  derived from two independent analyses and data sets match
 well. We discuss this in more detail in Sect. 6.

\subsection{Comparison of FIRES to other works at  z$>$1}

Papovich et al. (2005) have measured the evolution of the sizes in
the B--band restframe for galaxies in the HDF--N using WFPC2 and NICMOS
imaging. Papovich et al. measured sizes using SExtractor and not accounting for the
PSF effect in their measurements.
At z$\sim$2.3 they find a mean value of 2.3$\pm$0.3 kpc for
M(B)$\leq$-20.0. For galaxies with M(V)$\leq$-21.5 at z$\sim$2.5 we have
2.0$\pm$0.2 kpc. In both cases the error represents the uncertainty on the
mean. 
The agreement is encouraging taking into account the
different image quality  and  methods used for retrieving the half--light
radii.

At even larger redshifts, analysis of 1$<$z$<$6 galaxies based on the optical
bands (and consequently, matching the UV rest--frame) show a strong decrease in
size at a given UV luminosity with increasing redshift. This decrease scales
with z as: (1+z)$^{-1.5}$ (Ferguson et al. 2004) or as (1+z)$^{-1}$ (Bouwens et
al. 2004). In agreement with these results, in the redshift range 1$<$z$<$3 the
sizes at a given V--band luminosity presented here are well described by
(1+z)$^{-0.8\pm0.3}$. Consequently, the shape of the evolution is similar in the
UV and in the V-band restframe at least in the above redshift range.

\subsection{Comparison with previous HDF--S FIRES results}

Trujillo et al. (2004) explored the size evolution of the  galaxies contained in
the HDF--S. Their results are summarized in their Table 2. It is interesting to
check whether our current results, obtained with a larger sample, agree with
this previous analysis. At z$\sim$2.5, for galaxies more luminous than
2$\times$10$^{10}$h$_{70}$$^{-2}$L$_\odot$ they found that sizes were
$\sim$3$\pm$1 ($\pm$1 $\sigma$) times smaller than today counterparts. For
galaxies more massive than 2$\times$10$^{10}$h$_{70}$$^{-2}$M$_\odot$, sizes
were $\sim$1.4$\pm$0.5 ($\pm$1 $\sigma$) times smaller than local galaxies of
the same stellar mass.

For our current full data set,  at z$\sim$2.5,  galaxies more luminous than
3$\times$10$^{10}$h$_{70}$$^{-2}$L$_\odot$ are 3.8$\pm$0.5 ($\pm$2 $\sigma$)
smaller, and galaxies more massive than
3$\times$10$^{10}$h$_{70}$$^{-2}$M$_\odot$ are 2.1$\pm$0.3 ($\pm$2 $\sigma$)
smaller than same objects today. These values are larger than those obtained in
the HDF--S subsample  but are consistent within the uncertainties.

\subsection{Analytical description of the size evolution}

To provide  an analytical description of the rest--frame size evolution of the
galaxies in the redshift range 0$<$z$<$3, we have fitted the observed size
evolution   at a given luminosity (L$_V$$\gtrsim$ 3.4$\times$ 10$^{10}$
h$_{70}$$^{-2}$L$_\odot$) and at a given stellar mass
(M$_\star$$\gtrsim$3$\times$10$^{10}$h$_{70}$$^{-2}$M$_\odot$) to two different
analytical functions: a) (1+z)$^{\alpha}$ and b) H$^{\alpha}$(z). The parameters
of the fits are obtained by minimizing the $\chi^2$ error statistic.  To avoid
confusion   with  lines draw from comparison with theoretical models  we do not
overplot these fits in Fig. \ref{gemsfires}. The results of our fits, however,
are shown in Table \ref{modelfit}. In the low--n case a  better fit is obtained
using the function H$^{\alpha}$(z).

\subsection{Opacity effect on attenuation and size measurements}

The estimation of the brightness and the size of the galaxies is affected by
the dust content. Using the model of Popescu et al. (2000), the effect of dust
on the luminosity (Tuffs et al. 2004) and on the scalelength measurement
(M\"ollenhoff et al. 2006; in preparation) has been quantified: a larger amount
of dust increases the attenuation and the observed size (in terms of
scalelength) of the objects. The observed size is larger because  the dust
 is more strongly concentrated towards the central region of the galaxies
and consequently the flux gradient is flattened.

The size evolution presented in this paper is measured in relation to the
observed (uncorrected for dust) size of the local galaxies, consequently if the
dust opacity were not to change with redshift the observed evolution  presented
in this paper  would remain unchanged. However, it is likely that the opacity
of the galaxies changes with redshift. 

At a fixed inclination, bulge--to--total ratio and restframe wavelength, the
degree of attenuation and the increase in the observed scalelength due to dust
can be parametrized by the change in the central face--on optical depth. The
optical depth is a very uncertain quantity (even in the nearby universe) and
this makes a detailed evaluation of the effect of dust beyond the scope of this
paper. Consequently, we have not made any attempt to correct our results for
the effect of opacity. Nevertheless, in order to provide a crude estimation of
how a significant increase in opacity could affect our results we have made the
following exercise: let's assume a mean inclination of 30$^{\circ}$ and a
increase in the total central face--on optical depth in B--band from 4
(present--day galaxies) to 8 (high--z galaxies). This change implies a
transition  from an intermediate to a moderately optically thick case. In this
case, for a disk--like galaxy observed in the V--band restframe, the
attenuation increases by $\sim$0.2 mag (Tuffs et al. 2004; their Fig. 3 and
Table 4) and the scalength increases by $\sim$15\% (M\"ollenhoff et al. 2006;
in preparation). If we account for these numbers, the galaxies in our high--z
sample would be intrinsically brighter by $\sim$20\% and intrinsically smaller
by $\sim$15\%. In this sense, the observed (uncorrected for dust) size
evolution  presented in this paper would be  a lower limit of the actual size
evolution. If the opacity were smaller in the past then the situation would be
reversed, with our current estimation of the size evolution being an upper
limit.

\section{DISCUSSION}

We have greatly expanded the FIRES sample of galaxy rest--frame optical size
measurements, compared to Trujillo et al. (2004), and have combined these with
data from GEMS and SDSS. This combined data set allows us to analyze the
evolution of the luminosity--size and the stellar mass--size relations  for 
luminous (L$_V$$\gtrsim$ 3.4$\times$ 10$^{10}$ h$_{70}$$^{-2}$L$_\odot$) and
massive (M$_\star$$\gtrsim$3$\times$10$^{10}$h$_{70}$$^{-2}$M$_\odot$) galaxies
over 80\% of the Universe's age (0$<$z$<$3). During that time their
luminosity--size relation has changed strongly but the stellar mass--size
relation has  evolved  less than the luminosity--size relation. As
suggested in Trujillo et al. (2004) these two results can be reconciled when we
take into account the strong mass--to--light ratio evolution that galaxies have
experienced in the past. Such M/L evolution must also play a big role in
explaining the strong L$_{UV}$--r$_{e,UV}$ evolution seen in high--z samples (e.g.
Ferguson et al. 2003).

\begin{figure*}

\centering
\epsscale{.9}
\plotone{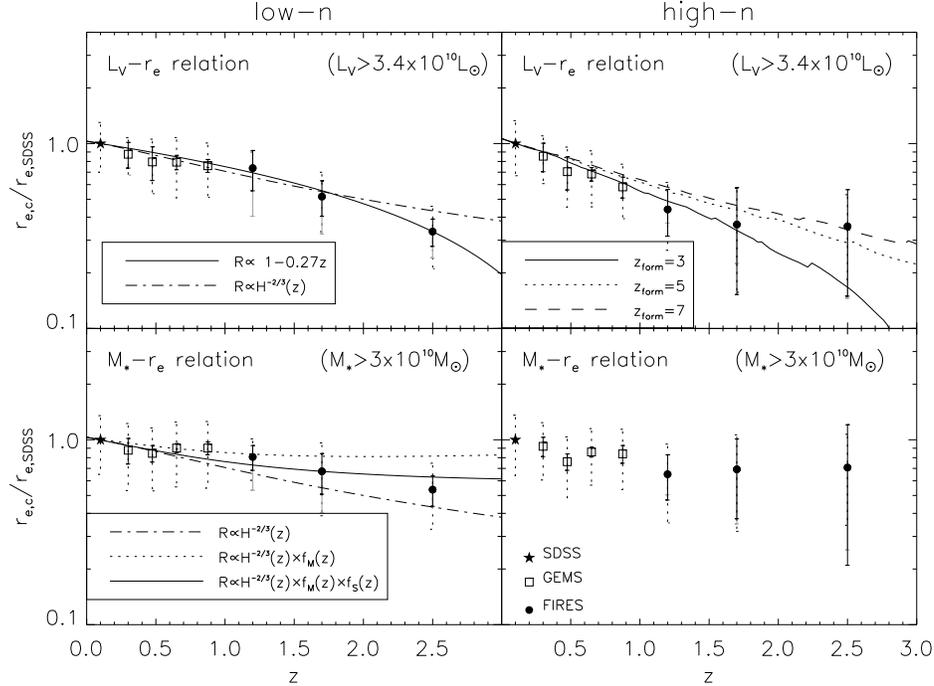}
\epsscale{1}

\caption{Redshift evolution of the ratio between the observed size and the
present--day mean size at a given luminosity (upper panels), and the analogous
ratio at a given mass (lower panels). The present--day values are derived from
the  SDSS sample (Shen et al. 2003). The comparison is restricted to  the
luminous (L$_V$$\gtrsim$3.4$\times$10$^{10}$h$_{70}$$^{-2}$L$_\odot$) and to
massive (M$_\star$$\gtrsim$3$\times$10$^{10}$h$_{70}$$^{-2}$M$_\odot$) galaxies.
Open squares correspond to the GEMS sample (McIntosh et al. 2005; Barden et al.
2005) for galaxies with z$<$1 and solid points indicate the results from FIRES.
The star indicates our local reference values from SDSS (mean $z$$\sim$0.1). We
present the dispersion (dashed error bars) and the uncertainty (2 $\sigma$) at
the mean determination (solid error bars) estimated from the
$\log$(r$_{e,c}$/r$_{e,SDSS}$) distribution. Grey error bars show how the
contribution of the small galaxies could affect the estimation of the mean.
$Left$ $column$: The dashed lines illustrate the expected evolution (Mo et al.
1998) at a fixed at fixed halo mass R$\propto$H$^{-2/3}$(z) normalized to be 1
at z=0.1. The predicted size evolution at a given luminosity for Milky Way type
objects (from the Bouwens \& Silk 2002 infall model) is indicated with a solid
line in the upper left panel. In the lower left panel we show (dotted line) the
Mo et al. (1998) size evolution at a given halo mass corrected by the evolution
of the stellar to halo mass f$_M$(z)=($M_{halo}/M_{\star}$)$^{1/3}$(z). The
solid line accounts for the transformation of the gas settled in the disk  into
stars by multypling the above correction for an extra factor
f$_S$(z)=$R_{\star}/R_{disk}$(z). $Right$ $column$. The different lines
illustrate the expected size evolution if the local luminosity--size relation
for early--type galaxies is evolved in luminosity as expected  for  single--age
stellar population models with different formation redshift (computed assuming a
Salpeter 1955 IMF using the PEGASE (Fioc \& Rocca--Volmerange 1997) code).}

\label{gemsfires}

\end{figure*}

Beyond the empirical result, it is of interest to compare  the observed evolution
with the theoretical predictions. In Fig. \ref{gemsfires} we show the expectations
from  semianalytical hierarchical and infall models for disk--like galaxies compared
to the observed size evolution. We first concentrate our attention on the evolution
of the sizes at a given luminosity. The semi--analytic hierarchical Mo et al. (1998)
model makes predictions on the disk size evolution at a given halo mass or circular
velocity, assuming that the disk mass is a fixed fraction of the halo mass. If one
then identifies Mo et al. disk mass with the stellar mass, or even the stellar
luminosity (as done e.g. by Ferguson et al. 2003) then a size--luminosity scaling of
H$^{-2/3}$(z) results. This scaling is shown in the top left panel of Fig.
\ref{gemsfires}, tantalizingly following the observations (except for the last point
at z=2.5). Yet, it must be borne in mind that this match implies a mean stellar M/L
that is constant with redshift, known to be incompatible with the color evolution of
the same galaxies. The agreement between H$^{-2/3}$(z) and the data must therefore be
considered fortuitous, rather than a direct confirmation of the Mo et al. model.

The infall (Bouwens \& Silk 2002) model  predicts directly the evolution of the
size at a given luminosity for Milky Way type objects. For that reason, we
compare the infall model only with the observed size evolution at a given
luminosity for galaxies with exponential--type profiles (upper left panel in
Fig. \ref{gemsfires}). We see that the agreement of this model with the observed
evolution is excellent for galaxies at all z. The  infall model, however, must
fail at higher z. In fact, this model shows an improbably fast decrease for
galaxies with z$>$2.5 and, for z$\gtrsim$3.7, this model produces sizes with
values less than zero.

If we focus now on the size evolution at a given disk mass and assume that the
stellar mass is a good indicator of the total baryonic mass settled in the disk
(which  the gas fraction at high redshift might invalidate), we can make a
comparison between the Mo et al.  model prediction and the observed size
evolution at a given disk mass. The bottom left panel of Fig.\ref{gemsfires} 
shows  that this hierarchical model (under the assumption stated in the
Introduction) produces a stronger evolution in the sizes than is observed.
However, at all z the model can not be rejected at 3$\sigma$ confidence level.
Consequently, although the observed evolution is weaker than the predicted size
evolution  R$\propto$ H$^{-2/3}$(z) at a fixed halo mass, this model can not be
rejected with the present dataset.

The Mo et al. (1998) model describes the evolution of the baryonic disk size at
a given halo mass whereas the data show  the stellar disk size evolution at a
given stellar mass. We now explore whether this difference maybe responsible for
the data model discrepancy apparent in the bottom left panel of
Fig.\ref{gemsfires}. We consider two aspects:  a) the ratio of the stellar mass
to the halo mass, M$_{\star}$/M$_{halo}$, can evolve with  redshift and b) the
ratio  of the stellar disk to the baryonic disk size,  R$_{\star}$/R$_{disk}$,
can also change.

These factors can be visualized by writing out the following identity:
\begin{equation}
\frac{R_{\star}}{M_{\star}^{1/3}}(z)=\frac{R_{disk}}{M_{halo}^{1/3}}(z) \times
\left(\frac{M_{halo}}{M_{\star}}(z)\right)^{1/3} \times
\frac{R_{\star}}{R_{disk}}(z) \end{equation}  where $R_{\star}/M_{\star}^{1/3}$
are the observables and  $R_{disk}/M_{halo}^{1/3}$ are the quantities more
inmediately predicted by Mo et al. (1998).

One possible choice to describe the accumulation of stellar mass within halos is
by the globally measured build--up of stellar mass:
M$_{\star}$/M$_{halo}$(z)$\sim$$<$$\rho_{\star}$(z)$>$, where we take
$<$$\rho_{\star}$(z)$>$ from Rudnick et al. (2003). Taking
R$_{\star}$/R$_{disk}$$\equiv$1 for now, this picture would predict a nearly
redshift--independent R$_{\star}$--M$_{\star}$ relation (dotted line in bottom
left panel of Fig. \ref{gemsfires}). However, this picture would imply that
stellar disks form from early--on in large halos and that the stellar disk,
already in its infancy (M$_{\star}$/M$_{halo}$$\ll$M$_{\star}$/M$_{halo}$(z=0))
samples the full angular momentum distribution of its large halo.

From a variety of observational and theoretical arguments R$_{\star}$/R$_{disk}$
cannot be unity at all epochs. As the solid line in Fig. \ref{gemsfires}
illustrates, through altering this assumption by 15--30\% (i.e. by assuming
$R_{\star}/R_{disk}(z)$$\propto$ H$^{-1/5}$(z)) it would be easy to match the
observations.

The  degree of evolution in the observed stellar mass--size relation with redshift
implies that galaxies must evolve  with time, increasing their size as they build up
their stellar mass.  Consequently, galaxies on average appear to grow inside--out.
Newly formed stars must preferentially reside at larger and larger radii (Trujillo \&
Pohlen 2005). 

In interpreting the evolution of spheroid--like objects  a different
reference hypothesis suggests itself: we analyze whether the decrease in typical
galaxy effective radius with lookback time at a given luminosity  is consistent
with a passively fading galaxy population.


To test the above idea  we plot on Fig. \ref{gemsfires} different tracks showing the
expected size evolution of a fading galaxy population with different formation
redshifts. These tracks are evaluated under the assumption that the shape of the
local  luminosity--size relation  does not change with redshift but for a shift of
the  relation to brighter luminosities at increasing z. The increase in the
luminosity with z is estimated by using the expected luminosity evolution from a
single burst at high--z (in our case, we have used z$_{form}$=3, 5 and 7) using the
P\'EGASE code (Fioc \& Rocca--Volmerange 1997). Following the same procedure as with
actual data, after shifting the luminosity--size relation we measure  the ratio
between the effective radii at a given luminosity for luminosities brighter than
3.4$\times$10$^{10}$h$_{70}$$^{-2}$L$_\odot$. From the comparison, we see that the
evolution of the luminosity-size relation for high--n galaxies is consistent
with a fading population of galaxies formed since z$\sim$3 to 7.


However, although the above agreement is encouraging,  the full population of
spheroid galaxies we see today is unlikely to be evolving passively since
z$\sim$3. The passive scenario is  against the observed evolution of the
co--moving total stellar mass density in passive red--sequence galaxies. This
density is lower at earlier epochs, amounting to a factor of $\sim$2 buildup
since z$\sim$1 (Chen et al. 2003; Bell et al. 2004; Cross et al. 2004) or a
factor of $\sim$10 since z$\sim$3 (Labb\'e et al. 2005). This change can not be
understood within a pure passive evolution scheme and it is in agreement with
the merger scenario proposed by Kauffmann \& Haehnelt (2000). In addition, 
Daddi et al. (2005) find 4 very compact (r$_e$$\lesssim$1 kpc) and massive
(M$_\star$$\gtrsim$10$^{11}$h$_{70}$$^{-2}$M$_\odot$) objects at z$\sim$1.7 in
the UDF. These objects could be the same class of compact galaxies that we find
here and could be found it at redshift as low as z$\sim$1 (see Fig. 9 from
McIntosh et al. 2005). In a $\Lambda$-CDM universe, Khochfar \& Silk (2006a)
find that  early-type galaxies at high redshifts merge from progenitors that
have  more cold gas available than their counter parts at low redshift. As a 
consequence they claim that the remnant should be smaller in size at high 
redshift (Khochfar \& Silk 2006b). These high--z spheroid--like objects are very
massive so it is not expected that their masses can increase dramatically since
then. So, we must expect a mechanism of growing in size very rapidly at
increasing their masses. As stated in the Introduction, the merger of
early--type galaxies could increase their sizes. If this is the case, repeated
mergers of the most massive spheroid-like objects that we observe at z$>$1.5
could bring them into the local observed stellar mass--size relation of
early--type galaxies. A more detailed analysis of the nature of these compact
objects in the FIRES sample will be presented in Toft et al. (2006) and Zirm et
al. (2006).

\begin{figure*}

\centering

\plotone{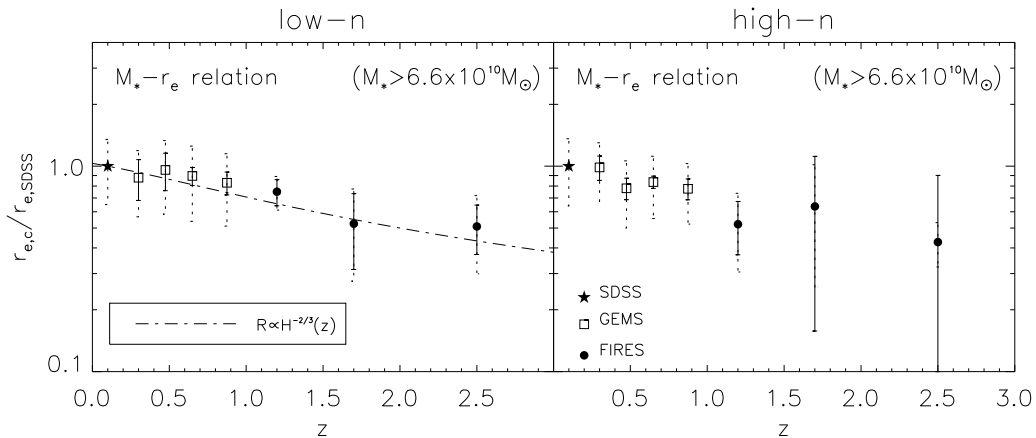}

\vspace{-4cm}

\caption{The ratio between observed size and expected size and at a given mass
from the local SDSS sample (Shen et al. 2003) as a function of z for  galaxies
more massive than our completeness mass limit
(M$_\star$$\gtrsim$6.6$\times$10$^{10}$h$_{70}$$^{-2}$M$_\odot$). The meaning
of the symbols is the same than in Fig. \ref{gemsfires}.}

\label{gemsfiresmassive}

\end{figure*}

We want to add a final cautionary note on the interpretation of the evolution of
the luminosity--size and stellar mass--size relations. There is a hint that the
degree of evolution of these relations could be different depending of the
luminosity and stellar mass range (or size) analyzed (Barden et al. 2005;
McIntosh et al. 2005). To test this we show in Fig. \ref{gemsfiresmassive} the
size evolution for galaxies more massive than our completeness mass limit
(M$_\star$$\gtrsim$6.6$\times$10$^{10}$h$_{70}$$^{-2}$M$_\odot$). In this case, 
the evolution in the sizes (at a given stellar mass) seems to be larger than if
we maintain the current limit. However, the uncertainty particularly at the
high--n sample is very large to make any strong conclusion.


\section{SUMMARY}

Using very deep near--infrared images of the HDF--S and the MS1054--03 field
from the FIRES survey we have analyzed the evolution of the luminosity--size
and stellar mass--size relation, measured  in their optical rest--frame, for
luminous (L$_V$$\gtrsim$3.4$\times$10$^{10}$h$_{70}$$^{-2}$L$_\odot$) and
massive (M$_\star$$\gtrsim$3$\times$10$^{10}$h$_{70}$$^{-2}$M$_\odot$) galaxies
with z$>$1. By combining HDF--S with the MS1054--03 field we have tripled the
number of galaxies with z$>$1 used in Trujillo et al. (2004). 

Several tests have been run in order to estimate the robustness of our
structural parameter estimates. From these tests we estimate an uncertainty in
our sizes of $\sim$25\% and in the concentration (S\'ersic index $n$) parameter
of $\sim$60\%. Moreover, we have briefly investigated whether our sample is
affected by surface brightness selection effects. As shown in that cursory
analysis, our magnitude selection criterion appear sufficiently conservative
enough to avoid such a concern.

Combining the analysis of FIRES data with the results obtained by  GEMS at 
z$<$1 (Barden et al. 2005; McIntosh et al. 2005) and tying both to the
present--day results from SDSS (Shen et al. 2003) we trace a detailed picture of
the evolution of the luminosity and stellar mass--size relations in the last
$\sim$11Gyrs. For less concentrated (low--n) objects,  at a given luminosity,
the typical sizes of the galaxies were $\sim$3 smaller at z$\sim$2.5 than those
we see today. In contrast, the stellar mass--size relation has evolved  less: we
see very little evolution  to z$\sim$1.2 and  a factor of $\sim$2 decrease in
size at a given stellar mass at z$\sim$2.5. The evolution at a given stellar
mass has  evolved proportional to (1+z)$^{-0.40\pm0.06}$. As pointed out by
Trujillo et al. (2004) the different evolution in the luminosity--size and the
stellar mass--size relation is explained by the fact that the M/L ratios of
high--z galaxies are lower than nowadays (or, the stellar populations were much
younger at earlier times). The   evolution observed in the stellar mass--size
relation combined with the fact that galaxies are producing new stars implies  
an inside--out growth of the galactic mass.

The observed luminosity--size relation evolution out to z$\sim$2.5 for low--n
objects matches very well the expected evolution for Milky--Way type objects
from infall models. For disk--like  galaxies, the semi--analytical hierarchical
predictions based on simple scaling relations between halos and baryons seem to
overestimate the observed  evolution of the stellar mass--size relation. The
discrepancy is in the sense that the observed galaxies at high redshift are
larger than expected from the model scalings. However, this model can not be
totally rejected with the current dataset.

 For highly concentrated (high--n) objects, the evolution of the
luminosity--size relation is consistent with (but does not necessarily imply)
pure luminosity evolution of a fading galaxy population. The evolution of the
sizes at a given stellar mass is proportional to (1+z)$^{-0.45\pm0.10}$.


\acknowledgements

We are happy to thank Shiyin Shen for providing us with the Sloan Digital Sky
Survey data used in this paper, E. F. Bell, E. Daddi and C. Heymans for useful
discussions. We would like to thank C. Moellenhoff, C.C. Popescu and R.J. Tuffs
for providing results from their calculations on the effects of dust on
measured scalelengths, prior to publication. We thank the staff at ESO for the
assistance in obtaining the FIRES data and  the Lorentz Center for its
hospitality and support. We thank the anonymous referee for the detailed
revision of our paper. Her/his comments have helped to improve the quality of
the manuscript.

Funding for the creation and distribution of the SDSS Archive has been provided
by the Alfred P. Sloan Foundation, the Participating Institutions, the National
Aeronautics and Space Administration, the National Science Foundation, the US
Department of Energy, the Japanese Monbukagakusho, and the Max-Planck Society.
The SDSS Web site is http://www.sdss.org. The SDSS is managed by the
Astrophysical Research Consortium (ARC) for the Participating Institutions. The
Participating Institutions are the University of Chicago, Fermilab, the
Institute for Advanced Study, the Japan Participation Group, Johns Hopkins
University, Los Alamos National Laboratory, the Max-Planck-Institut f\"ur
Astronomie (MPIA), the Max-Planck-Institut f\"ur Astrophysik (MPA), New Mexico
State University, University of Pittsburgh, Princeton University, the US Naval
Observatory, and the University of Washington.

GR acknowledges the support of a Goldberg fellowship at the National Optical
Astronomy Observatory (NOAO), which is operated by the Association of
Universities for Research in Astronomy (AURA), Inc., under a cooperative
agreement with the National Science Foundation.  GR also acknowledges the
financial support of the Sonderforschungsbereich 375 Astroteilchenphysik. MB
acknowledgs support from the {\it Verbundforschung} of the BMBF. DHM
acknowledges support from the National Aeronautics and Space Administration
(NASA) under LTSA Grant NAG5-13102 issued through the Office of Space Science.

\clearpage

\begin{deluxetable}{ccc}

\tablecaption{Moffat PSF fit to the sample images}

\tablewidth{0pc}

\tablehead{\colhead{Filter} & \colhead{$\beta$} & \colhead{FWHM}}

\startdata

& HDF--S &\\

\hline

J$_s$ & 3 & 0$''$.46 \\
H     & 3 & 0$''$.49 \\
K$_s$ & 3 & 0$''$.47 \\

\hline

& MS 1054--03 &\\

\hline

J$_s$ & 3.5 & 0$''$.48 \\
H     & 3   & 0$''$.46 \\
K$_s$ & 3   & 0$''$.53 \\

\enddata

\tablecomments{Col. (1): Filters used. Col. (2) and Col. (3)
$\beta$ and FWHM values estimated by fitting a Moffat PSF to star
profiles in the NIR images.}

\label{psfdata}

\end{deluxetable}

\begin{deluxetable}{ccccccc}

\tablecaption{Analytical descriptions of the results of our structural parameter
simulations}

\tablewidth{0pc}

\tablehead{\colhead{p$_{re}$} & \colhead{q$_{re}$} & \colhead{p$_n$}  
& \colhead{q$_n$} & \colhead{p$_{m}$}  & \colhead{K$_s$} & \colhead{n$_{input}$}}

\startdata


1.01 & 1.00 & 1.01 & 0.95 & 0.01  & 20--21  & $<$2.5 \\
0.95 & 0.97 & 0.98 & 0.81 & 0.01  & 21--22  & $<$2.5 \\
0.84 & 0.87 & 0.89 & 0.65 & 0.03  & 22--23  & $<$2.5 \\
0.90 & 0.94 & 1.06 & 0.90 & 0.04  & 20--21  & $>$2.5 \\
0.60 & 0.76 & 1.03 & 0.68 & 0.12  & 21--22  & $>$2.5 \\
0.55 & 0.71 & 0.67 & 0.74 & 0.16  & 22--23  & $>$2.5 \\

\enddata

\tablecomments{Cols. (1)--(5): Values of the parameters used in the analytical
fits to describe the difference between the input and the output r$_e$ and n in
our simulations. Col. (6) K$_s$ band magnitude bin. Col. (7) Value of the
input  index n.}

\label{corrections}

\end{deluxetable}

\clearpage

\begin{deluxetable}{ccccccccc}

\tablecaption{Properties of the MS1054--03 sample galaxies}

\tablewidth{0pc}

\tablehead{
\colhead{Galaxy} &  \colhead{K$_{s,tot}$} & \colhead{a$_e$} & \colhead{n}
  & \colhead{$\epsilon$} & \colhead{L$_V$(10$^{10}$ h$_{70}^{-2}$ 
L$_\odot$)}   &
\colhead{M(10$^{10}$ h$_{70}^{-2}$ M$_\odot$)}  & \colhead{$z$} & 
\colhead{Filter}}

\startdata

        1258 & 20.48 &  0.17 &  2.18 &  0.56 &   4.34 &  21.47 & 1.020 &  J$_s$ \\
         355 & 21.76 &  0.76 &  1.02 &  0.48 &   1.37 &   2.93 & 1.020 &  J$_s$ \\
        1638 & 22.64 &  0.17 &  3.06 &  0.48 &   0.54 &   1.56 & 1.040 &  J$_s$ \\
         848 & 22.01 &  0.56 &  0.52 &  0.76 &   1.13 &   1.77 & 1.040 &  J$_s$ \\
        1055 & 22.59 &  0.34 &  1.52 &  0.06 &   0.60 &   1.83 & 1.060 &  J$_s$ \\
        1132 & 20.87 &  0.53 &  1.32 &  0.56 &   2.51 &   9.03 & 1.060 &  J$_s$ \\
        1434 & 21.93 &  0.28 &  3.45 &  0.38 &   1.56 &   2.84 & 1.060 &  J$_s$ \\
        1566 & 21.78 &  0.48 &  0.92 &  0.80 &   1.02 &   3.66 & 1.060 &  J$_s$ \\
        1575 & 22.41 &  0.26 &  0.71 &  0.72 &   1.53 &   1.09 & 1.060 &  J$_s$ \\
        1801 & 21.36 &  0.12 &  3.15 &  0.17 &   2.44 &   7.71 & 1.070 &  J$_s$ \\
         830 & 22.29 &  0.31 &  1.05 &  0.52 &   1.51 &   1.23 & 1.073 &  J$_s$ \\
        1401 & 20.41 &  0.40 &  1.51 &  0.36 &   8.68 &   4.52 & 1.075 &  J$_s$ \\
         714 & 20.70 &  0.57 &  1.08 &  0.56 &   3.21 &   7.39 & 1.076 &  J$_s$ \\
        1229 & 22.78 &  0.33 &  0.83 &  0.24 &   1.37 &   1.37 & 1.080 &  J$_s$ \\
        1497 & 22.65 &  0.53 &  1.33 &  0.75 &   1.43 &   1.14 & 1.080 &  J$_s$ \\
         178 & 22.25 &  0.41 &  0.99 &  0.73 &   2.10 &   1.21 & 1.080 &  J$_s$ \\
         862 & 22.53 &  0.97 &  0.03 &  0.72 &   0.62 &   1.31 & 1.080 &  J$_s$ \\
         617 & 20.68 &  1.12 &  2.40 &  0.54 &   4.43 &  19.59 & 1.100 &  J$_s$ \\
        1216 & 21.29 &  0.12 &  3.96 &  0.50 &   2.70 &  15.27 & 1.120 &  J$_s$ \\
         147 & 22.33 &  0.35 &  0.57 &  0.53 &   1.38 &   2.39 & 1.120 &  J$_s$ \\
         150 & 22.55 &  0.20 &  1.00 &  0.48 &   1.54 &   2.55 & 1.120 &  J$_s$ \\
        1768 & 21.28 &  0.54 &  1.31 &  0.56 &   2.50 &   7.41 & 1.120 &  J$_s$ \\
         359 & 22.59 &  0.72 &  0.81 &  0.64 &   1.38 &   1.07 & 1.120 &  J$_s$ \\
         100 & 21.13 &  0.15 &  3.54 &  0.51 &   3.09 &  15.69 & 1.140 &  J$_s$ \\
        1172 & 22.50 &  0.14 &  4.04 &  0.36 &   0.91 &   6.68 & 1.140 &  J$_s$ \\
         460 & 22.31 &  0.52 &  0.05 &  0.59 &   2.16 &   1.14 & 1.140 &  J$_s$ \\
         527 & 21.02 &  0.46 &  5.31 &  0.28 &   3.36 &  16.33 & 1.140 &  J$_s$ \\
         749 & 21.15 &  0.30 &  6.00 &  0.39 &   3.74 &   9.72 & 1.140 &  J$_s$ \\
        1440 & 22.18 &  0.55 &  1.17 &  0.49 &   2.44 &   2.75 & 1.160 &  J$_s$ \\
        1785 & 20.05 &  0.33 &  2.94 &  0.88 &  11.31 &  32.58 & 1.170 &  J$_s$ \\
         494 & 21.87 &  0.20 &  6.24 &  0.38 &   2.44 &   3.15 & 1.175 &  J$_s$ \\
        1273 & 22.15 &  0.14 &  3.35 &  0.26 &   2.12 &   1.82 & 1.180 &  J$_s$ \\
         481 & 21.81 &  0.38 &  4.42 &  0.46 &   1.89 &   7.74 & 1.180 &  J$_s$ \\
        1535 & 21.75 &  0.66 &  0.50 &  0.20 &   1.77 &   0.98 & 1.182 &  J$_s$ \\
         508 & 21.49 &  1.10 &  0.46 &  0.87 &   1.61 &   5.76 & 1.189 &  J$_s$ \\
        1301 & 21.91 &  0.15 &  2.88 &  0.26 &   1.83 &  30.94 & 1.200 &  J$_s$ \\
         161 & 20.44 &  0.26 &  6.68 &  0.74 &   7.38 &   7.31 & 1.200 &  J$_s$ \\
        1786 & 21.46 &  0.15 &  4.00 &  0.23 &   2.82 &  25.32 & 1.200 &  J$_s$ \\
        1621 & 21.85 &  0.63 &  0.15 &  0.62 &   2.50 &   4.93 & 1.220 &  J$_s$ \\
         306 & 20.90 &  0.77 &  3.29 &  0.23 &   6.66 &   5.47 & 1.220 &  J$_s$ \\
          45 & 22.36 &  0.59 &  0.96 &  0.41 &   2.07 &   1.73 & 1.220 &  J$_s$ \\
         614 & 20.75 &  0.37 &  1.78 &  0.37 &   5.26 &  11.28 & 1.220 &  J$_s$ \\
         441 & 20.52 &  0.50 &  3.52 &  0.60 &   6.88 &  17.65 & 1.230 &  J$_s$ \\
        1176 & 22.88 &  0.23 &  2.97 &  0.12 &   1.61 &   0.85 & 1.234 &  J$_s$ \\
         743 & 22.48 &  0.13 &  4.45 &  0.41 &   1.11 &   5.88 & 1.240 &  J$_s$ \\
         774 & 21.97 &  0.63 &  0.79 &  0.32 &   3.03 &   3.36 & 1.240 &  J$_s$ \\
        1474 & 21.93 &  0.74 &  1.02 &  0.21 &   3.72 &   2.56 & 1.245 &  J$_s$ \\
        1267 & 22.45 &  0.96 &  6.08 &  0.34 &   0.92 &   0.53 & 1.246 &  J$_s$ \\
        1438 & 21.71 &  0.53 &  0.85 &  0.55 &   3.23 &   3.43 & 1.247 &  J$_s$ \\
        1266 & 22.34 &  0.35 &  0.67 &  0.66 &   1.64 &   2.95 & 1.280 &  J$_s$ \\
        1280 & 22.05 &  0.07 &  4.38 &  0.40 &   2.95 &   3.75 & 1.280 &  J$_s$ \\
         737 & 21.09 &  0.59 &  1.54 &  0.69 &   6.09 &   9.31 & 1.280 &  J$_s$ \\
        1226 & 22.87 &  0.41 &  0.70 &  0.67 &   1.64 &   0.79 & 1.295 &  J$_s$ \\
        1256 & 20.50 &  0.46 &  1.55 &  0.14 &  10.30 &  22.85 & 1.300 &  J$_s$ \\
        1637 & 21.77 &  1.01 &  0.96 &  0.85 &   1.89 &   3.52 & 1.300 &  J$_s$ \\
         487 & 22.51 &  2.12 &  0.38 &  0.85 &   1.10 &   1.77 & 1.300 &  J$_s$ \\
          54 & 21.78 &  0.65 &  0.94 &  0.61 &   3.60 &   3.94 & 1.300 &  J$_s$ \\
         869 & 22.51 &  0.25 &  1.03 &  0.75 &   1.84 &   1.96 & 1.300 &  J$_s$ \\
         971 & 22.98 &  0.20 &  1.16 &  0.91 &   0.93 &   1.52 & 1.300 &  J$_s$ \\
        1071 & 21.52 &  0.18 &  4.30 &  0.51 &   3.50 &  16.62 & 1.320 &  J$_s$ \\
        1456 & 21.58 &  0.13 &  6.00 &  0.54 &   5.36 &   2.77 & 1.320 &  J$_s$ \\
         438 & 22.17 &  0.46 &  1.00 &  0.26 &   3.31 &   2.03 & 1.320 &  J$_s$ \\
          67 & 21.06 &  0.55 &  3.35 &  0.23 &   7.06 &   6.20 & 1.326 &  J$_s$ \\
        1120 & 22.68 &  0.67 &  0.38 &  0.47 &   2.09 &   1.02 & 1.340 &  J$_s$ \\
        1218 & 22.21 &  1.10 &  0.10 &  0.81 &   3.17 &   2.54 & 1.340 &  J$_s$ \\
         479 & 22.60 &  0.78 &  1.31 &  0.76 &   1.61 &   1.86 & 1.340 &  J$_s$ \\
         732 & 22.61 &  0.55 &  0.96 &  0.27 &   1.94 &   1.12 & 1.360 &  J$_s$ \\
         795 & 21.68 &  0.58 &  0.46 &  0.56 &   3.01 &   8.46 & 1.360 &  J$_s$ \\
         845 & 21.79 &  0.53 &  0.56 &  0.13 &   4.46 &   2.88 & 1.360 &  J$_s$ \\
        1719 & 20.79 &  0.40 &  2.10 &  0.21 &   7.73 &  18.70 & 1.400 &  J$_s$ \\
        1763 & 22.17 &  0.27 &  1.89 &  0.36 &   2.90 &   4.03 & 1.400 &  J$_s$ \\
        1781 & 21.21 &  0.18 &  4.00 &  0.20 &   4.95 &  28.18 & 1.400 &  J$_s$ \\
        1249 & 22.53 &  1.05 &  0.02 &  0.75 &   2.71 &   2.26 & 1.420 &  J$_s$ \\
         379 & 22.80 &  0.73 &  0.47 &  0.76 &   1.91 &   1.42 & 1.420 &  J$_s$ \\
         552 & 22.32 &  0.60 &  0.66 &  0.74 &   2.48 &   3.09 & 1.420 &  J$_s$ \\
          40 & 22.20 &  0.91 &  1.40 &  0.54 &   1.85 &   4.36 & 1.440 &  J$_s$ \\
        1341 & 22.51 &  0.33 &  0.39 &  0.26 &   2.30 &   2.80 & 1.460 &  J$_s$ \\
        1792 & 22.10 &  1.19 &  1.09 &  0.49 &   2.28 &   8.61 & 1.460 &  J$_s$ \\
         259 & 22.93 &  0.31 &  0.78 &  0.42 &   1.96 &   1.17 & 1.460 &  J$_s$ \\
         831 & 22.17 &  1.09 &  0.32 &  0.68 &   4.78 &   4.54 & 1.460 &  J$_s$ \\
         878 & 22.46 &  0.17 &  1.58 &  0.25 &   3.18 &   1.64 & 1.460 &  J$_s$ \\
        1378 & 22.90 &  0.36 &  0.72 &  0.58 &   2.12 &   1.06 & 1.480 &  J$_s$ \\
         706 & 22.81 &  0.61 &  0.02 &  0.38 &   1.46 &   1.55 & 1.480 &  J$_s$ \\
        1292 & 22.84 &  0.23 &  2.04 &  0.24 &   2.44 &   2.00 & 1.500 &  J$_s$ \\
        1671 & 21.28 &  0.48 &  1.98 &  0.11 &   7.02 &  10.69 & 1.520 &  H \\
         999 & 22.69 &  0.12 &  2.30 &  0.95 &   1.50 &   5.43 & 1.520 &  H \\
        1268 & 22.80 &  0.55 &  0.71 &  0.59 &   2.57 &   2.55 & 1.540 &  H \\
        1591 & 22.15 &  0.28 &  1.61 &  0.37 &   4.69 &   4.06 & 1.540 &  H \\
         321 & 22.20 &  0.52 &  5.23 &  0.23 &   3.12 &   3.93 & 1.540 &  H \\
        1155 & 21.87 &  0.36 &  0.70 &  0.46 &   4.07 &  10.08 & 1.560 &  H \\
        1124 & 22.61 &  0.62 &  1.28 &  0.73 &   2.65 &   3.47 & 1.580 &  H \\
        1540 & 21.75 &  0.18 &  1.51 &  0.96 &   4.61 &  14.09 & 1.600 &  H \\
        1704 & 21.50 &  1.36 &  2.31 &  0.77 &   8.03 &   7.45 & 1.600 &  H \\
        1774 & 21.60 &  0.08 &  4.27 &  0.92 &   5.66 &  25.11 & 1.600 &  H \\
        1797 & 21.82 &  0.17 &  2.09 &  0.86 &   4.06 &  28.02 & 1.600 &  H \\
          37 & 21.06 &  0.24 &  0.77 &  0.54 &  10.28 &  24.12 & 1.600 &  H \\
         807 & 22.99 &  0.32 &  2.19 &  0.19 &   2.58 &   2.68 & 1.620 &  H \\
         928 & 21.19 &  0.22 &  3.83 &  0.80 &   9.05 &  31.61 & 1.620 &  H \\
         110 & 22.84 &  0.38 &  6.45 &  0.56 &   2.48 &   1.66 & 1.640 &  H \\
        1199 & 22.39 &  0.56 &  4.13 &  0.71 &   3.57 &   2.55 & 1.640 &  H \\
        1586 & 21.89 &  0.29 &  1.38 &  0.43 &   5.36 &   7.26 & 1.640 &  H \\
        1753 & 22.86 &  0.33 &  5.65 &  0.75 &   1.70 &   3.96 & 1.640 &  H \\
         281 & 22.63 &  0.57 &  0.75 &  0.57 &   3.88 &   2.62 & 1.640 &  H \\
         582 & 22.92 &  0.19 &  2.40 &  0.42 &   2.82 &   1.98 & 1.640 &  H \\
         962 & 22.16 &  0.15 &  4.14 &  0.58 &   3.63 &   5.58 & 1.640 &  H \\
         157 & 22.86 &  0.44 &  0.42 &  0.49 &   2.19 &   1.52 & 1.660 &  H \\
        1577 & 22.30 &  0.14 &  1.52 &  0.18 &   4.29 &   6.05 & 1.660 &  H \\
        1695 & 21.98 &  1.47 &  2.97 &  0.54 &   4.51 &   8.77 & 1.660 &  H \\
        1776 & 22.03 &  0.48 &  0.35 &  0.38 &   4.75 &   8.12 & 1.660 &  H \\
        1350 & 22.01 &  0.62 &  1.25 &  0.39 &   4.84 &   3.81 & 1.700 &  H \\
         402 & 22.62 &  0.31 &  5.19 &  0.58 &   3.68 &   1.88 & 1.700 &  H \\
         523 & 22.33 &  0.24 &  0.46 &  0.52 &   3.06 &  12.27 & 1.700 &  H \\
         528 & 20.32 &  0.31 &  2.84 &  0.56 &  24.52 &  39.42 & 1.700 &  H \\
         561 & 20.39 &  0.33 &  2.12 &  0.82 &  21.70 &  38.67 & 1.700 &  H \\
         713 & 20.46 &  0.27 &  1.76 &  0.57 &  20.55 &  38.83 & 1.700 &  H \\
        1295 & 21.55 &  0.28 &  2.98 &  0.34 &   7.21 &  17.20 & 1.720 &  H \\
        1459 & 21.32 &  0.27 &  2.17 &  0.37 &   7.44 &  21.36 & 1.740 &  H \\
        1309 & 22.90 &  0.26 &  0.99 &  0.58 &   2.16 &   8.66 & 1.800 &  H \\
        1650 & 22.96 &  0.30 &  1.71 &  0.46 &   2.94 &   3.04 & 1.820 &  H \\
          73 & 21.80 &  0.06 &  4.00 &  0.72 &   6.14 &  25.10 & 1.820 &  H \\
        1717 & 22.62 &  0.12 &  3.35 &  0.64 &   3.47 &   6.67 & 1.860 &  H \\
           7 & 22.44 &  0.29 &  0.37 &  0.55 &   3.62 &  24.71 & 1.860 &  H \\
        1714 & 21.47 &  0.69 &  0.92 &  0.73 &   8.97 &  23.27 & 1.880 &  H \\
         723 & 21.32 &  0.81 &  1.20 &  0.57 &   9.78 &  52.80 & 1.880 &  H \\
         842 & 22.78 &  0.25 &  0.79 &  0.27 &   4.11 &   3.57 & 1.900 &  H \\
        1530 & 22.73 &  0.37 &  0.19 &  0.17 &   4.13 &   2.71 & 1.920 &  H \\
         386 & 22.88 &  0.34 &  1.15 &  0.62 &   2.18 &   5.98 & 1.920 &  H \\
        1702 & 22.56 &  1.33 &  0.38 &  0.79 &   4.57 &   2.43 & 1.940 &  H \\
        1373 & 22.07 &  0.55 &  0.91 &  0.19 &   6.69 &   5.73 & 1.960 &  H \\
        1335 & 22.97 &  0.41 &  1.97 &  0.53 &   2.51 &  13.07 & 1.980 &  H \\
         926 & 22.18 &  0.84 &  0.32 &  0.73 &   4.44 &  17.94 & 1.980 &  H \\
        1294 & 22.51 &  0.49 &  0.83 &  0.51 &   5.12 &   6.87 & 2.000 &  H \\
        1457 & 22.85 &  0.49 &  0.38 &  0.41 &   2.85 &   3.50 & 2.000 &  H \\
        1571 & 21.78 &  0.29 &  0.41 &  0.29 &   9.23 &   6.27 & 2.020 &  H \\
        1061 & 21.48 &  0.31 &  2.11 &  0.20 &  11.54 &  62.24 & 2.120 &  H \\
        1265 & 22.06 &  0.13 &  4.21 &  0.20 &   9.49 &  13.55 & 2.140 &  H \\
        1550 & 22.81 &  0.33 &  1.38 &  0.52 &   4.11 &   5.15 & 2.140 &  H \\
         325 & 22.77 &  0.45 &  0.86 &  0.60 &   4.78 &   2.42 & 2.140 &  H \\
         914 & 21.67 &  0.05 &  6.73 &  0.55 &  12.82 &  28.78 & 2.160 &  H \\
         383 & 21.28 &  0.78 &  0.08 &  0.57 &  18.80 &  17.25 & 2.180 &  H \\
         852 & 22.80 &  0.52 &  0.63 &  0.65 &   3.85 &  10.00 & 2.180 &  H \\
        1644 & 22.71 &  0.33 &  1.00 &  0.00 &   4.27 &   9.16 & 2.200 &  H \\
         972 & 22.62 &  0.44 &  0.63 &  0.90 &   5.07 &   4.94 & 2.200 &  H \\
        1144 & 22.56 &  0.03 &  1.02 &  0.00 &   4.77 &  23.61 & 2.220 &  H \\
        1612 & 22.01 &  0.46 &  0.88 &  0.70 &   8.93 &  38.79 & 2.240 &  H \\
         847 & 22.18 &  0.32 &  0.47 &  0.36 &   6.74 &  26.59 & 2.240 &  H \\
        1538 & 22.98 &  0.72 &  1.00 &  0.34 &   3.80 &   8.27 & 2.300 &  H \\
         846 & 22.40 &  0.50 &  0.66 &  0.59 &   7.53 &   9.95 & 2.320 &  H \\
        1086 & 22.56 &  0.17 &  1.00 &  0.62 &   7.11 &   3.38 & 2.340 &  H \\
        1410 & 22.17 &  0.07 &  3.64 &  0.11 &  11.24 &  10.65 & 2.400 &  H \\
        1035 & 21.43 &  0.23 &  6.51 &  0.42 &  19.64 &  13.72 & 2.425 &  H \\
        1356 & 21.53 &  0.04 &  4.73 &  0.54 &  18.51 &  25.47 & 2.427 &  H \\
        1383 & 21.41 &  0.66 &  1.62 &  0.44 &  18.45 &  61.06 & 2.430 &  H \\
        1547 & 22.51 &  0.16 &  0.28 &  0.22 &   8.16 &  13.51 & 2.480 &  H \\
        1656 & 22.40 &  0.20 &  1.00 &  0.50 &  10.04 &   2.65 & 2.500 &  H \\
         645 & 21.52 &  0.80 &  2.54 &  0.73 &  22.61 &  24.38 & 2.520 &  H \\
        1239 & 22.70 &  0.26 &  0.63 &  0.60 &   8.90 &   3.96 & 2.620 &  K$_s$ \\
        1496 & 22.93 &  0.08 &  3.00 &  0.30 &   7.87 &   4.65 & 2.700 &  K$_s$ \\
        1237 & 22.97 &  0.10 &  1.50 &  0.00 &   7.27 &   9.71 & 2.760 &  K$_s$ \\
         773 & 22.85 &  0.19 &  5.00 &  0.75 &   8.72 &   2.97 & 2.800 &  K$_s$ \\
        1353 & 22.75 &  0.53 &  0.95 &  0.46 &   9.62 &  11.77 & 2.840 &  K$_s$ \\
        1100 & 22.50 &  1.03 &  1.13 &  0.44 &  12.34 &  18.51 & 2.880 &  K$_s$ \\
        1253 & 22.05 &  0.68 &  0.77 &  0.19 &  21.36 &  18.78 & 3.000 &  K$_s$ \\
        1041 & 22.72 &  0.21 &  0.50 &  0.34 &  14.10 &  11.44 & 3.480 &  K$_s$ \\
        1666 & 22.78 &  0.76 &  0.22 &  0.49 &  13.86 &  10.83 & 3.540 &  K$_s$ \\
        1211 & 22.82 &  0.62 &  0.39 &  0.59 &  15.85 &   8.26 & 3.780 &  K$_s$ \\
         543 & 22.84 &  0.33 &  0.91 &  0.26 &  24.25 &   9.39 & 4.760 &  K$_s$ \\
         549 & 22.27 &  0.28 &  1.42 &  0.40 &  58.94 &  29.30 & 4.900 &  K$_s$ \\
         472 & 22.65 &  0.16 &  1.81 &  0.07 &  38.90 &  17.49 & 4.960 &  K$_s$ \\
         859 & 22.50 &  0.28 &  1.67 &  0.42 &  61.64 &  31.61 & 5.200 &  K$_s$ \\
        1690 & 22.77 &  0.88 &  1.41 &  0.56 & 135.91 & 257.24 & 5.400 &  K$_s$ \\
        1758 & 22.11 &  0.31 &  0.84 &  0.73 & 287.54 & 450.87 & 5.740 &  K$_s$ \\
        1467 & 22.87 &  0.05 &  9.02 &  0.56 &  51.67 &  27.30 & 5.960 &  K$_s$ \\
        1778 & 22.72 &  0.17 &  1.99 &  0.22 &  73.47 &  45.17 & 6.000 &  K$_s$ \\
           6 & 22.52 &  0.94 &  0.97 &  0.74 & 320.90 & 637.74 & 6.000 &  K$_s$ \\

\enddata

\tablecomments{Col. (1): Catalog identification numbers (see F\"orster
Schreiber et al. 2005).  Col (2): $K_s$--band total magnitudes. Col. (3):
Semimajor axis optical restframe half--light radii (arcsec). The typical
uncertainty on the size determination is 25\%. Col. (4): S\'ersic index n . The
typical uncertainty on the shape determination is 50\%. Col (5): intrinsic
(i.e. the recovered non--seeing affected) ellipticity. Col (6): Rest--frame
V--band luminosity. The typical uncertainty on the luminosity determination is
30\%. Col (7): Stellar mass. Col (8): Redshift Col (9): 
Filter used to measure the size of the galaxies}

\label{tabdata}

\end{deluxetable}

\clearpage

\clearpage

\begin{deluxetable}{ccccccccc}

\tablecaption{Properties of the HDF-S sample galaxies}

\tablewidth{0pc}

\tablehead{
\colhead{Galaxy} &  \colhead{K$_{s,tot}$} & \colhead{a$_e$} & \colhead{n}
  & \colhead{$\epsilon$} & \colhead{L$_V$(10$^{10}$ h$_{70}^{-2}$ 
L$_\odot$)}   &
\colhead{M(10$^{10}$ h$_{70}^{-2}$ M$_\odot$)}  & \colhead{$z$} & 
\colhead{Filter}}

\startdata

  224 & 21.83 &  0.25 &  1.27 &  0.25 &   1.16 &   2.22 & 1.020 &  J$_s$ \\
  753 & 22.90 &  0.23 &  1.02 &  0.26 &   0.84 &   0.60 & 1.020 &  J$_s$ \\
10008 & 22.33 &  0.15 &  2.13 &  0.36 &   0.66 &   2.16 & 1.040 &  J$_s$ \\
  152 & 23.00 &  0.37 &  0.84 &  0.17 &   0.88 &   0.64 & 1.060 &  J$_s$ \\
  241 & 21.72 &  0.74 &  1.19 &  0.56 &   1.51 &   3.20 & 1.060 &  J$_s$ \\
   79 & 21.49 &  0.56 &  0.70 &  0.53 &   2.53 &   2.83 & 1.080 &  J$_s$ \\
   18 & 21.20 &  0.31 &  1.12 &  0.36 &   2.24 &   6.71 & 1.100 &  J$_s$ \\
  249 & 22.60 &  0.78 &  1.81 &  0.67 &   0.60 &   1.29 & 1.100 &  J$_s$ \\
  565 & 20.75 &  0.48 &  0.87 &  0.28 &   4.72 &   5.98 & 1.114 &  J$_s$ \\
  686 & 21.06 &  0.32 &  1.61 &  0.03 &   3.21 &   5.77 & 1.116 &  J$_s$ \\
  493 & 20.97 &  0.36 &  4.57 &  0.55 &   3.29 &   4.14 & 1.120 &  J$_s$ \\
   45 & 20.89 &  0.18 &  3.19 &  0.09 &   4.16 &   8.34 & 1.140 &  J$_s$ \\
  206 & 22.71 &  0.37 &  0.48 &  0.37 &   1.37 &   0.68 & 1.152 &  J$_s$ \\
  276 & 20.89 &  0.23 &  1.95 &  0.63 &   4.10 &  12.52 & 1.160 &  J$_s$ \\
  644 & 22.67 &  0.22 &  0.85 &  0.18 &   0.83 &   4.58 & 1.160 &  J$_s$ \\
  669 & 23.27 &  0.47 &  0.35 &  0.07 &   0.95 &   0.47 & 1.200 &  J$_s$ \\
  404 & 22.75 &  0.49 &  0.55 &  0.27 &   1.33 &   1.22 & 1.220 &  J$_s$ \\
   27 & 20.22 &  0.48 &  3.21 &  0.17 &   8.68 &  16.44 & 1.230 &  J$_s$ \\
  251 & 22.79 &  0.67 &  0.61 &  0.76 &   1.11 &   1.45 & 1.240 &  J$_s$ \\
  254 & 20.31 &  0.22 &  3.11 &  0.04 &  10.13 &  15.94 & 1.270 &  J$_s$ \\
  101 & 22.23 &  0.37 &  2.16 &  0.63 &   2.48 &   2.94 & 1.280 &  J$_s$ \\
  149 & 23.18 &  0.24 &  0.75 &  0.41 &   0.61 &   1.53 & 1.280 &  J$_s$ \\
  470 & 20.39 &  0.49 &  0.84 &  0.14 &   8.03 &  12.52 & 1.284 &  J$_s$ \\
  502 & 23.20 &  0.84 &  0.91 &  0.70 &   0.69 &   0.96 & 1.300 &  J$_s$ \\
  771 & 22.86 &  0.25 &  0.46 &  0.33 &   0.92 &   1.41 & 1.300 &  J$_s$ \\
  145 & 22.35 &  0.65 &  7.00 &  0.51 &   1.53 &   2.05 & 1.320 &  J$_s$ \\
  395 & 22.65 &  0.25 &  0.56 &  0.15 &   1.84 &   1.75 & 1.320 &  J$_s$ \\
  637 & 21.95 &  0.35 &  3.42 &  0.36 &   3.43 &   3.71 & 1.320 &  J$_s$ \\
  199 & 21.68 &  0.27 &  2.90 &  0.25 &   2.64 &  12.80 & 1.340 &  J$_s$ \\
  791 & 22.98 &  0.39 &  0.74 &  0.47 &   1.19 &   1.20 & 1.360 &  J$_s$ \\
  437 & 23.16 &  0.68 &  1.05 &  0.71 &   1.19 &   1.19 & 1.380 &  J$_s$ \\
  201 & 22.96 &  0.36 &  0.71 &  0.47 &   2.03 &   1.32 & 1.400 &  J$_s$ \\
  408 & 23.09 &  0.16 &  1.68 &  0.59 &   1.54 &   1.15 & 1.400 &  J$_s$ \\
  785 & 21.57 &  0.54 &  0.42 &  0.43 &   4.42 &   6.81 & 1.400 &  J$_s$ \\
  751 & 23.13 &  0.19 &  0.97 &  0.14 &   1.41 &   1.65 & 1.420 &  J$_s$ \\
  302 & 21.55 &  0.65 &  0.83 &  0.21 &   6.00 &   6.87 & 1.439 &  J$_s$ \\
10001 & 21.54 &  0.27 &  1.36 &  0.14 &   5.12 &   6.45 & 1.440 &  J$_s$ \\
   61 & 23.03 &  0.78 &  3.42 &  0.45 &   1.20 &   1.39 & 1.440 &  J$_s$ \\
  783 & 22.51 &  0.27 &  0.60 &  0.33 &   1.77 &   2.49 & 1.440 &  J$_s$ \\
  781 & 22.73 &  0.77 &  1.10 &  0.66 &   2.21 &   1.53 & 1.480 &  J$_s$ \\
  620 & 22.16 &  0.25 &  1.42 &  0.30 &   4.64 &   3.04 & 1.558 &  H \\
  628 & 22.37 &  0.15 &  0.08 &  0.40 &   2.36 &   6.49 & 1.580 &  H \\
  675 & 22.23 &  0.37 &  0.30 &  0.33 &   3.29 &   4.44 & 1.600 &  H \\
  724 & 23.35 &  0.32 &  1.03 &  0.62 &   1.15 &   1.55 & 1.620 &  H \\
  583 & 22.90 &  0.10 &  1.01 &  0.15 &   1.80 &   8.90 & 1.640 &  H \\
  349 & 23.17 &  0.65 &  2.43 &  0.25 &   2.18 &   1.06 & 1.680 &  H \\
  233 & 23.38 &  0.07 &  1.00 &  0.10 &   2.10 &   1.45 & 1.720 &  H \\
  754 & 23.16 &  0.25 &  6.00 &  0.46 &   1.50 &   3.68 & 1.760 &  H \\
  267 & 21.84 &  0.69 &  0.51 &  0.37 &   7.00 &   6.93 & 1.820 &  H \\
  810 & 22.80 &  0.10 &  3.06 &  0.44 &   2.40 &   5.50 & 1.920 &  H \\
  600 & 22.33 &  0.67 &  4.87 &  0.49 &   6.01 &   7.14 & 1.960 &  H \\
  500 & 23.25 &  0.24 &  0.50 &  0.73 &   1.83 &   3.91 & 2.020 &  H \\
  290 & 21.95 &  0.23 &  0.62 &  0.31 &   9.51 &   5.19 & 2.025 &  H \\
  257 & 22.10 &  0.71 &  0.76 &  0.35 &   7.66 &   4.65 & 2.027 &  H \\
   21 & 23.49 &  0.66 &  7.26 &  0.94 &   2.35 &   0.81 & 2.040 &  H \\
   96 & 23.35 &  0.29 &  1.08 &  0.28 &   2.88 &   1.16 & 2.060 &  H \\
  776 & 22.44 &  0.22 &  1.64 &  0.26 &   5.76 &   4.01 & 2.077 &  H \\
  173 & 23.23 &  0.31 &  0.38 &  0.48 &   2.89 &   1.76 & 2.140 &  H \\
  496 & 22.40 &  0.27 &  0.86 &  0.40 &   4.91 &   9.17 & 2.140 &  H \\
  729 & 22.73 &  0.47 &  1.93 &  0.63 &   5.14 &   1.87 & 2.140 &  H \\
  143 & 23.37 &  0.49 &  0.24 &  0.50 &   2.89 &   1.93 & 2.160 &  H \\
  242 & 23.43 &  0.38 &  0.74 &  0.85 &   2.57 &   1.13 & 2.160 &  H \\
  219 & 23.35 &  0.44 &  0.80 &  0.81 &   2.60 &   2.76 & 2.200 &  H \\
  375 & 22.80 &  0.55 &  0.05 &  0.64 &   4.12 &   6.42 & 2.240 &  H \\
  767 & 22.54 &  0.12 &  6.00 &  0.38 &   6.25 &  20.77 & 2.300 &  H \\
  161 & 23.42 &  0.06 &  6.00 &  0.58 &   2.66 &  11.35 & 2.340 &  H \\
  595 & 23.48 &  0.30 &  0.32 &  0.32 &   3.05 &   1.92 & 2.400 &  H \\
  176 & 22.93 &  1.06 &  1.85 &  0.36 &   5.70 &   8.39 & 2.500 &  H \\
  363 & 22.42 &  0.60 &  1.09 &  0.46 &   9.65 &   4.09 & 2.500 &  H \\
10006 & 23.32 &  0.10 &  4.90 &  0.18 &   4.97 &   2.88 & 2.652 &  K$_s$ \\
  656 & 22.70 &  0.33 &  1.10 &  0.32 &   8.60 &  31.14 & 2.740 &  K$_s$ \\
  452 & 22.84 &  0.44 &  0.36 &  0.55 &   8.45 &   6.25 & 2.760 &  K$_s$ \\
  806 & 22.67 &  0.17 &  4.15 &  0.68 &  10.04 &   3.60 & 2.789 &  K$_s$ \\
  807 & 22.70 &  0.28 &  0.87 &  0.30 &   9.93 &   3.80 & 2.790 &  K$_s$ \\
  657 & 22.53 &  0.70 &  0.25 &  0.16 &  12.14 &   7.18 & 2.793 &  K$_s$ \\
  294 & 23.34 &  0.45 &  0.36 &  0.35 &   5.74 &   3.99 & 2.820 &  K$_s$ \\
  453 & 23.28 &  0.15 &  4.43 &  0.51 &   6.11 &  16.63 & 2.900 &  K$_s$ \\
  494 & 23.00 &  0.72 &  1.78 &  0.47 &   9.14 &   4.67 & 3.000 &  K$_s$ \\
  534 & 22.78 &  0.32 &  0.96 &  0.46 &  10.93 &   9.23 & 3.000 &  K$_s$ \\
  465 & 23.38 &  0.35 &  0.50 &  0.50 &   6.39 &   6.68 & 3.040 &  K$_s$ \\
  397 & 23.42 &  0.40 &  0.33 &  0.71 &   6.47 &  12.07 & 3.080 &  K$_s$ \\
  448 & 23.46 &  0.19 &  0.88 &  0.33 &   5.67 &   2.56 & 3.140 &  K$_s$ \\
  622 & 23.08 &  0.40 &  0.25 &  0.55 &   7.97 &   3.98 & 3.140 &  K$_s$ \\
  624 & 23.19 &  0.22 &  0.50 &  0.25 &   8.61 &  18.40 & 3.160 &  K$_s$ \\
   98 & 23.09 &  0.19 &  0.31 &  0.28 &   9.50 &  19.73 & 3.160 &  K$_s$ \\
  813 & 23.32 &  0.40 &  1.69 &  0.47 &   6.42 &   2.85 & 3.240 &  K$_s$ \\
   80 & 22.72 &  0.45 &  1.27 &  0.22 &  18.13 &   7.59 & 3.840 &  K$_s$ \\

\enddata

\tablecomments{Col. (1): Catalog identification numbers (see F\"orster
Schreiber et al. 2005).  Col (2): $K_s$--band total magnitudes. Col. (3):
Semimajor axis optical restframe half--light radii (arcsec). The typical
uncertainty on the size determination is 25\%. Col. (4): S\'ersic index n . The
typical uncertainty on the shape determination is 50\%. Col (5): intrinsic
(i.e. the recovered non--seeing affected) ellipticity. Col (6): Rest--frame
V--band luminosity. The typical uncertainty on the luminosity determination is
30\%. Col (7): Stellar mass. Col (8): Redshift Col (9): 
Filter used to measure the size of the galaxies}

\label{tabdatasur}

\end{deluxetable}

\clearpage

\begin{deluxetable}{ccc}

\tablecaption{Mean size evolution vs redshift}

\tablewidth{0pc}

\tablehead{\colhead{$<$z$>$} & \colhead{low-n} & \colhead{high-n}}

\startdata

& L$_V$$\gtrsim$3.4$\times$10$^{10}$h$_{70}$$^{-2}$L$_\odot$ &\\

\hline

0.1  & 1 & 1 \\
0.3  & 0.88$\pm$0.13 &  0.85$\pm$0.15 \\
0.5  & 0.80$\pm$0.16 &  0.70$\pm$0.15 \\
0.65 & 0.79$\pm$0.07 &  0.68$\pm$0.06 \\
0.9  & 0.76$\pm$0.06 &  0.58$\pm$0.08 \\
1.2  & 0.74$\pm$0.18 &  0.44$\pm$0.12 \\
1.7  & 0.52$\pm$0.12 &  0.36$\pm$0.22 \\
2.5  & 0.33$\pm$0.06 &  0.37$\pm$0.20 \\
\hline

 & M$_\star$$\gtrsim$3$\times$10$^{10}$h$_{70}$$^{-2}$M$_\odot$ &\\

\hline

0.1  & 1 & 1 \\
0.3  & 0.88$\pm$0.14 &  0.92$\pm$0.11\\
0.5  & 0.84$\pm$0.09 &  0.76$\pm$0.08\\
0.65 & 0.90$\pm$0.05 &  0.86$\pm$0.06\\
0.9  & 0.90$\pm$0.07 &  0.84$\pm$0.10\\
1.2  & 0.81$\pm$0.13 &  0.65$\pm$0.18 \\
1.7  & 0.67$\pm$0.16 &  0.69$\pm$0.32 \\
2.5  & 0.54$\pm$0.10 &  0.71$\pm$0.50 \\
\enddata

\tablecomments{Col. (1): Mean redshift of the bin Col. (2) and Col. (3)
r$_e$(z)/r$_e$(0.1) and the 2$\sigma$ uncertainty on the mean values estimated from
the $\log$(r$_{e,c}$/r$_{e,SDSS}$) distribution.}

\label{sizeevoldata}

\end{deluxetable}

\clearpage

\begin{deluxetable}{ccc}

\tablecaption{Analytical Fits to the size evolution}

\tablewidth{0pc}
\tablehead{\colhead{Fit} & \colhead{$\alpha$} & \colhead{$\tilde{\chi}^2$}}

\startdata

& L$_V$$\gtrsim$3.4$\times$10$^{10}$h$_{70}$$^{-2}$L$_\odot$ (low-n) &\\

\hline


(1+z)$^{\alpha}$  & -0.84$\pm$0.05 & 2.29 \\
H$^{\alpha}$(z)   & -0.83$\pm$0.05 & 0.71 \\

\hline

& L$_V$$\gtrsim$3.4$\times$10$^{10}$h$_{70}$$^{-2}$L$_\odot$ (high-n) &\\

\hline


(1+z)$^{\alpha}$  & -1.01$\pm$0.08 & 0.25 \\
H$^{\alpha}$(z)   & -1.13$\pm$0.09 & 0.68 \\

\hline

 & M$_\star$$\gtrsim$3$\times$10$^{10}$h$_{70}$$^{-2}$M$_\odot$ (low-n) &\\

\hline


(1+z)$^{\alpha}$  & -0.40$\pm$0.06 & 0.89 \\
H$^{\alpha}$(z)   & -0.43$\pm$0.07 & 0.50 \\

\hline

 & M$_\star$$\gtrsim$3$\times$10$^{10}$h$_{70}$$^{-2}$M$_\odot$ (high-n) &\\

\hline


(1+z)$^{\alpha}$  & -0.45$\pm$0.10 & 0.59 \\
H$^{\alpha}$(z)   & -0.54$\pm$0.12 & 0.73 \\

\enddata

\tablecomments{Col. (1): Analytical expression used to fit the data Col. (2) Value
of the parameter measured including 1$\sigma$ error bar and 
Col. (3) Reduced $\tilde{\chi}^2$ value of the fit.}

\label{modelfit}

\end{deluxetable}

\clearpage

\clearpage
\thispagestyle{empty}
\setlength{\voffset}{-20mm}
\clearpage
\setlength{\voffset}{0mm}

\end{document}